\documentclass[journal,10pt]{IEEEtran}

\usepackage{stfloats}
\usepackage{amsmath}
\usepackage{amssymb,latexsym}
\usepackage{graphicx}
\usepackage{color,cite,times}
\usepackage{epstopdf}
\usepackage{algorithmic}
\usepackage[ruled,linesnumbered] {algorithm2e}
\usepackage{booktabs}
\usepackage{bbm} 
\usepackage{url}
\usepackage{fancyhdr}
\usepackage{cleveref}
\usepackage{verbatim}
\usepackage{multirow}
\usepackage{makecell}
\usepackage{graphicx}
\usepackage{indentfirst}
\usepackage{cases}
\usepackage{amsthm,amsmath}
\newcommand{\RNum}[1]{\uppercase\expandafter{\romannumeral #1\relax}}
\usepackage{colortbl}
\usepackage{extarrows}
\usepackage{threeparttable}
\usepackage{diagbox}
\usepackage{pifont}
\usepackage{cancel}
\usepackage{framed}
\usepackage{overpic}
\usepackage[most]{tcolorbox}
\usepackage{subfigure}
\usepackage[framemethod=tikz]{mdframed}
\usepackage{lipsum}

\newcommand{\bsm}{\boldsymbol}

\newcommand{\mbf}{\mathbf}
\newcommand{\diag}{\mathrm{diag}}

\newtheorem{remk}{Remark}

\newtheorem{prop}{Proposition}

\newtheorem{assump}{Assumption}

\definecolor{gray}{RGB}{192,192,192}
\definecolor{dred}{RGB}{192,0,0}

\DeclareMathOperator{\rank}{rank}

\begin{document}

\title{End-to-End Learning for Task-Oriented Semantic Communications Over MIMO Channels: An Information-Theoretic Framework}

	\author{Chang Cai, \IEEEmembership{Graduate Student Member, IEEE,}
		Xiaojun Yuan, \IEEEmembership{Senior Member, IEEE,} \\
		and Ying-Jun Angela Zhang, \IEEEmembership{Fellow, IEEE} 
			\thanks{Chang Cai and Ying-Jun Angela Zhang are with the Department of Information Engineering, The Chinese University of Hong Kong, Hong Kong (e-mail: cc021@ie.cuhk.edu.hk; yjzhang@ie.cuhk.edu.hk).
				
				Xiaojun Yuan is with
				the National Key Laboratory of Wireless Communications, 
				University of Electronic Science and Technology of China, Chengdu 611731, China (e-mail: xjyuan@uestc.edu.cn). 
			}
		}
	\maketitle

\begin{abstract}
	This paper addresses the problem of end-to-end (E2E) design of learning and communication in a task-oriented semantic communication system.
	In particular, we consider a multi-device cooperative edge inference system over a wireless multiple-input multiple-output (MIMO) multiple access channel, where multiple devices transmit extracted features to a server to perform a classification task.
	We formulate the E2E design of feature encoding, MIMO precoding, and classification as a conditional mutual information maximization problem.
	However, it is notoriously difficult to design and train an E2E network that can be adaptive to both the task dataset and different channel realizations.
	Regarding network training, we propose a decoupled pretraining framework that separately trains the feature encoder and the MIMO precoder, with a maximum \textit{a posteriori} (MAP) classifier employed at the server to generate the inference result.
	The feature encoder is pretrained exclusively using the task dataset, while the MIMO precoder is pretrained solely based on the channel and noise distributions.
	Nevertheless, we manage to align the pretraining objectives of each individual component with the E2E learning objective, so as to approach the performance bound of E2E learning.
	By leveraging the decoupled pretraining results for initialization, the E2E learning can be conducted with minimal training overhead.
	Regarding network architecture design, we develop two deep unfolded precoding networks that effectively incorporate the domain knowledge of the solution to the decoupled precoding problem. 
	Simulation results on both the CIFAR-10 and ModelNet10 datasets verify that the proposed method achieves significantly higher classification accuracy compared to various baselines.
\end{abstract}


\section{Introduction}
Driven by the recent advances in artificial intelligence (AI), the next-generation wireless networks are foreseeable to support many emerging new services such as augmented reality, autonomous driving, and smart healthcare.
These applications often require frequent and massive data exchange, posing an unaffordable burden to current wireless systems with very limited radio spectrum availability.
On the other hand, the ultimate goal of communication in these applications is usually no longer the exact recovery of transmitted data, but the efficient execution of a certain task.
This gives rise to a new research topic named task-oriented communication \cite{gunduz2023beyond}, a.k.a. semantic communication \cite{kaibin2021jcin, dusit2023survey}, which transmits only the information essential for the successful execution of the task, thereby relieving the communication burden.

Task-oriented communications involve encoding design at the transmitter for task-relevant feature extraction, together with decoding design at the receiver to accomplish a specific task \cite{zhijin2021single, pingzhang2022jsac}.
The codecs are usually parameterized by neural networks (NNs) owing to their powerful representation and generalization capabilities.
We refer to the choice of the codecs as learning design.
Meanwhile, the multiple-input multiple-output (MIMO) technique, which benefits from high array gains of multiple antennas, is a crucial feature in current wireless protocols and is expected to support future task-oriented communication systems. 
The MIMO transceiver design therein should be revisited and revised considering the paradigm shift from accurate bit transmission to successful task completion.
In this regard, it is pivotal to design a holistic system where learning and MIMO communication are jointly considered in order to improve efficiency and reliability. 

Most existing works on task-oriented communications \cite{zhijin2021multiuser, zhang2023scan, wu2023deep}, however, directly reuse the MIMO transceivers developed for traditional communications.
These traditional communication designs aim at throughput maximization, mean-square error (MSE) minimization, bit error rate (BER) minimization, etc.
The misalignment between the objectives of MIMO communication and task execution may hinder the exploitation of the full benefits of task-oriented communications.
For example, ref. \cite{zhijin2021multiuser} applies a linear minimum mean-square error (LMMSE) detector to recover the transmitted features for image retrieval, machine translation, and visual question answering tasks.
Refs. \cite{zhang2023scan, wu2023deep} employ the singular value decomposition (SVD) based precoding for an image transmission task.
Although the SVD based precoding is capacity-achieving for communication over a MIMO Gaussian channel, it is not necessarily optimal in terms of the end-to-end (E2E) task execution performance.

To our best knowledge, only a few initial attempts \cite{cai2023multi, GM2023Wen} consider to align the communication objective with successful task execution, focusing specifically on a classification task.
The transceivers therein are expected to promote class-wise separability on the received/recovered features, thereby improving classification accuracy.
This problem fundamentally differs from that in traditional communications, in that any forms of distortion are tolerable as long as they do not deteriorate the classification accuracy.
Due to the absence of analytical forms of classification accuracy, 
refs. \cite{cai2023multi, GM2023Wen} adopt maximal coding rate reduction (MCR$^2$) \cite{yu2020learning} and discriminant gain as surrogate accuracy measures, respectively.
These metrics rely on heuristics to characterize the separability of different classes of received/recovered features, serving as the optimization objectives of precoding \cite{cai2023multi} or receive beamforming \cite{GM2023Wen}.
Despite achieving noticeable accuracy gain, these methods are not deemed optimal since they artificially separate the design of wireless communication and learning based codecs, lacking the flexibility for E2E fine-tuning. 
In fact, the communication and learning aspects are inherently coupled in task-oriented communications, 
which motivates the need for E2E communication-learning co-design to bridge the potential performance gap \cite{aylin2023wcm}.

In this paper, we study a multi-device edge inference system over a MIMO multiple access channel, where multiple devices transmit low-dimensional features to a server to perform a classification task.
We formulate the joint design of feature encoding, MIMO precoding, and classification as an E2E learning problem. 
This formulation presents two main challenges.
From the network training aspect, the E2E network should learn the parameters based on not only the task dataset (e.g., multi-view image and label pairs), but also the entire distributions of wireless channel and noise.
This incurs a huge training overhead due to the need for simultaneous sampling over these datasets/distributions. 
The high dimensional channel matrix in a MIMO setting makes the E2E learning even more complicated and inefficient.
From the network design aspect, the network architecture should be carefully crafted to capture the intrinsic structures of the problem.
Although some heavily engineered networks (such as ResNet \cite{he2016deep} and ViT \cite{ViT2021ICLR}) have achieved empirical success for feature encoding and classification in machine learning literature, the network architecture suitable for precoding in the considered task-oriented communication system is not well understood yet.

To tackle the aforementioned challenges, we propose a decoupled design framework built upon the original E2E formulation, eliminating the need for sampling simultaneously from the task dataset and the channel distribution.
The decoupled design framework, on one hand, can serve as the pretraining method to individually train the feature encoder and the MIMO precoder prior to E2E learning, which effectively reduces the E2E training overhead. 
On the other hand, this framework can also serve as the guiding principle for precoding network construction. 
We propose to use the deep unfolding technique \cite{hershey2014deep, Yonina2019SPM}, which unfolds the iterations in the decoupled precoding optimization algorithm into a layer-wise structure. 
The unfolded structure introduces a set of learnable parameters to improve the performance.
Meanwhile, it preserves the domain knowledge of the decoupled precoding problem, which is more reliable and interpretable than a black-box NN designed by trial and error.
The main contributions of this paper are summarized as follows.
\begin{itemize}
	\item \textbf{Information-theoretic E2E learning formulation:}
	We formulate the joint design of feature encoding, MIMO precoding, and classification as an E2E conditional mutual information maximization problem.
	This formulation aims to preserve the maximum amount of target label information in the received features, conditioned on the channel state.
	It non-trivially extends the works in \cite{jiawei2021task, jiawei2023multidevice, songjie2023digital, infomax2023sensors} by incorporating an explicit characterization of the wireless channel into formulation.
	\item \textbf{Decoupled pretraining framework of E2E learning:}
	We establish a decoupled pretraining framework based on the original E2E formulation. 
	The feature encoder is pretrained exclusively using the task dataset, while the precoder is pretrained solely based on the channel and noise distributions.
	We employ a maximum \textit{a posteriori} (MAP) classifier at the server to generate the inference result.
	We manage to align the pretraining objectives of each individual component with the E2E learning objective.
	By doing so, the decoupled pretraining serves as an appropriate initialization for E2E learning, which drastically reduces the E2E training overhead.
	Moreover, the decoupled pretraining framework establishes a close connection between mutual information and coding rate reduction \cite{yu2020learning}, providing an information-theoretic understanding on the heuristic communication-learning separation in \cite{cai2023multi}.
	\item \textbf{Deep unfolding based precoding network design:}
	We first propose a deep unfolded precoding network, referred to as vanilla DU-BCA precoder, built upon the block coordinate ascent (BCA) algorithm \cite{cai2023multi} for solving the decoupled precoding problem.
	We identify the inherent limitations of this vanilla design, stemming from the inappropriate parameterization of matrix inversion and Lagrange multipliers. 
	To avoid these operations, we modify one block of the base algorithm to a majorization-minimization (MM) implementation.
	Accordingly, we develop an enhanced DU-BCA-MM precoder that refines the vanilla design.
\end{itemize}
Simulation results on the CIFAR-10 \cite{krizhevsky2009learning} and ModelNet10 \cite{wu20153d} datasets showcase the superior performance of the proposed E2E learning method compared to various baselines.
The performance gain comes from both the aligned communication-learning objective and the customized architecture of the precoding network.

The remainder of this paper is organized as follows.
In Section \ref{Sec_MDEI}, we establish the probabilistic model of multi-device edge inference and present a general formulation of the E2E learning problem.
In Section \ref{decoupled-design}, we propose a decoupled pretraining framework for feature encoding, precoding, and classification.
Section \ref{Sec-E2E-Learning} elaborates the E2E learning algorithm.
Section \ref{DU-BCA} provides a vanilla design of the precoding network based on deep unfolding.
Section \ref{DU-BCA-MM} further refines the vanilla design by enhancing both the base algorithm and the network architecture.
Section \ref{Sec_simulation} provides extensive simulation results to evaluate the effectiveness of the proposed method. 
Finally, we conclude this paper in Section \ref{Sec_Conclusions}.

\textit{Notations:}
We denote random variables by capital letters (e.g., $X$ and $Y$) and their realizations by lowercase letters (e.g., $\mbf{x}$ and $y$).
We abbreviate a sequence $(X_1, \dots, X_K)$ of $K$ random variables by $X_{1:K}$, and their realizations $(\mbf{x}_1, \dots, \mbf{x}_K)$ by $\mbf{x}_{1:K}$.
Matrices are denoted by uppercase boldface letters, e.g., $\mbf{A}$.
We use $\mbf{A}^{\sf T}$, $\mbf{A}^{\sf H}$, and $\mbf{A}^{-1}$ to denote the transpose, conjugate transpose, and inverse of matrix $\mbf{A}$, respectively.
Sets are denoted by calligraphic letters, e.g., $\mathcal{A}$.
Moreover, we use $\otimes$, $\diag\{ \cdot \}$, $\mathbb{E} [ \cdot ]$, and $\mathrm{tr} ( \cdot )$ to represent the Kronecker product, the diagonal operator, the expectation operator, and the trace of a square matrix, respectively.  
The probability density function (PDF) of a circularly symmetric complex Gaussian (CSCG) random vector $\mbf{x} \in \mathbb{C}^N$ with mean $\bsm{\mu}$ and covariance matrix $\bsm{\Sigma}$ is denoted by $\mathcal{CN} (\mbf{x}; \bsm{\mu}, \bsm{\Sigma}) = \exp \left(-(\mbf{x}-\bsm{\mu})^{\sf H}\bsm{\Sigma}^{-1}(\mbf{x}-\bsm{\mu})\right)/\left(\pi^N \det (\bsm{\Sigma}) \right)$.

\begin{figure*}
	[t]
	\centering
	\subfigure[Block diagram]
	{\includegraphics[width=1.3\columnwidth]{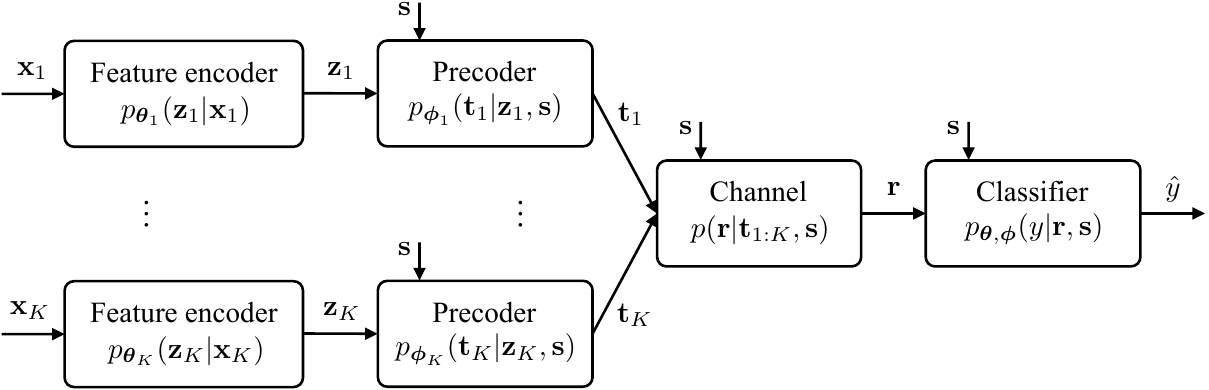}\label{Probabilistic}} \hfil
	\subfigure[Markov model]
	{\includegraphics[width=.45\columnwidth]{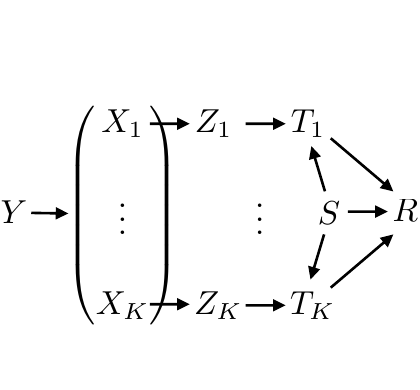}\label{Markov}} 
	\caption{Block diagram and Markov model of the considered multi-device edge inference system.}
	\label{Model}
\end{figure*}
\section{Wireless Multi-Device Edge Inference System} \label{Sec_MDEI}
We consider a multi-device edge inference system, where multiple distributed devices collaborate with an edge server to perform an inference task.
Due to limited communication resources, it is impracticable to directly transmit the raw data collected by each device (e.g., high-resolution images of a common object captured from different views) to the server.
Instead, each device aims to extract a compact representation of the collected data, named feature, for efficient transmission over a wireless link.
In this section, we first establish a probabilistic model of the multi-device edge inference system.
We put a special emphasis on the characterization of the wireless fading channel in such a system, which is often ignored or oversimplified in existing literature \cite{jiawei2021task, jiawei2023multidevice, songjie2023digital, infomax2023sensors, pingzhang2022jsac, mxtao2023jsac}.
This characterization introduces a new research problem of the dedicated transceiver design apart from the traditional focus on feature encoding and decoding.
\subsection{Probabilistic Modeling}
The multi-device edge inference system comprises $K$ edge devices and an edge server, as depicted in Fig. \ref{Probabilistic}.
Let $\mathcal{K} \triangleq \left\{1, \dots, K\right\}$.
The observations $(\mbf{x}_1, \dots, \mbf{x}_K)$ and its target $y$ (e.g., the category of an object) are deemed as the realization of the random variables $(X_1, \dots, X_K, Y)$ with joint distribution $p(\mbf{x}_1, \dots, \mbf{x}_K, y)$.
The observations could be distinct or redundant.
To perform cooperative inference, each device extracts the feature $\mbf{z}_k \in \mathbb{C}^{D_k}$ from its input $\mbf{x}_k$ through a probabilistic feature encoder $p_{\bsm{\theta}_k}(\mbf{z}_k |\mbf{x}_k)$ parameterized by $\bsm{\theta}_k$. 
We adopt a deterministic feature encoder $\mbf{z}_k = f_{\bsm{\theta}_k}(\mbf{x}_k)$ in this paper, which can be regarded as a special case of the probabilistic one by noting its equivalent form as 
\begin{align}
	p_{\bsm{\theta}_k}(\mbf{z}_k |\mbf{x}_k) = \delta (\mbf{z}_k -  f_{\bsm{\theta}_k}\left(\mbf{x}_k)\right),
\end{align}
where $\delta ( \cdot)$ is the Dirac delta function.

We consider linear analog modulation for feature transmission over a MIMO multiple access channel.
Let $\mbf{H}_k \in \mathbb{C}^{N_{\sf r} \times N_{\mathsf{t}, k}}$ be the baseband equivalent channel from device $k$ to the edge server, where $N_{{\mathsf t}, k}$ and $N_{\mathsf r}$ denote the numbers of antennas equipped at device $k$ and at the edge server, respectively.
The feature $\mbf{z}_k$ undergoes precoding before transmission. 
Each device employs a linear precoder $\mbf{V}_k \in \mathbb{C}^{N_{\mathsf{t}, k} \times D_k}$ to produce the transmitted signal vector $\mbf{t}_k \in \mathbb{C}^{N_{{\sf t}, k}}$ as
\begin{align}
	\mbf{t}_k = \mbf{V}_k \mbf{z}_k.
\end{align}
The received signal vector $\mbf{r} \in \mathbb{C}^{N_{\sf r}}$ is then given by
\begin{align}\label{sum-signal-model}
	\mbf{r} = \sum_{k \in \mathcal{K}}\mbf{H}_k \mbf{t}_k + \mbf{n},
\end{align}
where $\mbf{n} \sim \mathcal{CN} (\mbf{0}, \sigma^2 \mbf{I})$ is the additive white Gaussian noise (AWGN).
By defining $\mbf{H} \triangleq \left[\mbf{H}_1 , \dots, \mbf{H}_K\right]$, $\mbf{V} \triangleq \diag \left\{\mbf{V}_1, \dots, \mbf{V}_K\right\}$, $\mbf{z} \triangleq \left[\mbf{z}_1^{\sf H}, \dots, \mbf{z}_K^{\sf H}\right]^{\sf H}$, and $\mbf{t} \triangleq \left[\mbf{t}_1^{\sf H}, \dots, \mbf{t}_K^{\sf H}\right]^{\sf H}$, 
we rewrite \eqref{sum-signal-model} more compactly as
\begin{align}\label{signal-model}
	\mbf{r}  = \mbf{H} \mbf{V} \mbf{z} + \mbf{n}
	= \mbf{H} \mbf{t} + \mbf{n}.
\end{align}
\begin{remk}
	Since the feature dimension can often be greater than the effective channel rank, i.e., $D_k > \rank (\mbf{H}_k)$, it is necessary to allocate multiple time slots to transmit a feature vector for accurate inference. 
	The signal model developed above can be naturally extended to accommodate this case, allowing for joint management of precoding across the relevant time slots. 
	Assuming a total of $O$ time slots for transmission,
	we can extend the definition of the channel matrix $\mbf{H}_k$ by collecting these matrices for different time slots into a block diagonal matrix as 
	$\mbf{H}_k \triangleq \diag \big\{\mbf{H}_k (1) ,\dots, \mbf{H}_k (O) \big\} \in \mathbb{C}^{O N_{\mathsf r} \times O N_{{\mathsf t}, k}}$, where $\mbf{H}_k (o)$ stands for the channel at the $o$-th time slot.
	Similarly, the definitions of $\mbf{V}_k$, $\mbf{t}_k$, $\mbf{n}$, and $\mbf{r}$ can be extended as
	$\mbf{V}_k \triangleq \left[\mbf{V}_k^{\sf H}(1), \dots, \mbf{V}_k^{\sf H}(O) \right]^{\sf H} \in \mathbb{C}^{O N_{{\sf t}, k} \times D_k}$,
	$\mbf{t}_k \triangleq \left[\mbf{t}_k^{\sf H}(1), \dots, \mbf{t}_k^{\sf H}(O) \right]^{\sf H} \in \mathbb{C}^{O N_{{\sf t}, k}}$,
	$\mbf{n} \triangleq \left[\mbf{n}^{\sf H}(1), \dots, \mbf{n}^{\sf H}(O) \right]^{\sf H} \in \mathbb{C}^{O N_{{\sf r}}}$, and
	$\mbf{r} \triangleq \left[\mbf{r}^{\sf H} (1), \dots, \mbf{r}^{\sf H} (O)\right]^{\sf H} \in \mathbb{C}^{O N_{{\sf r}}}$, respectively.
	To simplify notation, we assume $O = 1$ throughout the derivation.
	The results of $O>1$ are provided in the experiments. 
\end{remk}

We are now ready to present the probabilistic modeling of wireless transmission.
We introduce the state $\mbf{s} \triangleq \{\mbf{H}, \sigma\}$ as the collection of the wireless channel and the noise level.
The precoding relies on $\mbf{s}$ and is determined by the function $\mbf{V}_k = g_{\bsm{\phi}_k}(\mbf{s})$ through parameters $\bsm{\phi}_k$, which can be expressed in a probabilistic form as
\begin{align} \label{precoding-probabilistic}
	p_{\bsm{\phi}_k}(\mbf{t}_k | \mbf{z}_k, \mbf{s}) = \delta(\mbf{t}_k - g_{\bsm{\phi}_k}(\mbf{s})\mbf{z}_k).
\end{align}
It is noteworthy that we do not assume the availability of the state $\mbf{s}$ at the devices.
Instead, the dependence on $\mbf{s}$ in \eqref{precoding-probabilistic} is achieved by calculating precoding matrices at the server and then feeding back to the devices.
The distribution that maps $(\mbf{t}_1, \dots, \mbf{t}_K)$ to $\mbf{r}$ with $\mbf{s}$ given is
\begin{align}
	p(\mbf{r}| \mbf{t}_{1:K}, \mbf{s}) = \mathcal{CN} \left(\mbf{H}\mbf{t}, \sigma^2\mbf{I}\right).
\end{align}

Fig. \ref{Markov} presents the corresponding Markov model, satisfying
\begin{align} \label{Markov-rule}
	p_{\bsm{\theta}, \bsm{\phi}}(\mbf{r}|\mbf{x}_{1:K}, \mbf{s}) = p(\mbf{r}|\mbf{t}_{1:K}, \mbf{s})
	p_{\bsm{\phi}} (\mbf{t}_{1:K}|\mbf{z}_{1:K}, \mbf{s})  
	p_{\bsm{\theta}}(\mbf{z}_{1:K}|\mbf{x}_{1:K}),
\end{align}
where $p_{\bsm{\phi}} (\mbf{t}_{1:K}|\mbf{z}_{1:K}, \mbf{s}) = \prod_{k \in \mathcal{K}} p_{\bsm{\phi}_k}(\mbf{t}_k|\mbf{z}_k, \mbf{s})$ with $\bsm{\phi} \triangleq \{\bsm{\phi}_k\}_{k \in \mathcal{K}}$, 
and $p_{\bsm{\theta}}(\mbf{z}_{1:K}|\mbf{x}_{1:K}) = \prod_{k \in \mathcal{K}} p_{\bsm{\theta}_k}(\mbf{z}_k |\mbf{x}_k)$ with $\bsm{\theta} \triangleq \{\bsm{\theta}_k\}_{k \in \mathcal{K}}$.
Theoretically, the optimal inference model is given by the posterior $p (y|\mbf{r}, \mbf{s})$ without the need to recover the transmitted features.
The posterior is fully determined by $\bsm{\theta}$ and $\bsm{\phi}$ by applying the Bayes' law: 
\begin{align}
	p_{\bsm{\theta}, \bsm{\phi}} (y|\mbf{r}, \mbf{s}) &= \frac{p_{\bsm{\theta}, \bsm{\phi}}(y, \mbf{r}|\mbf{s})}{p_{\bsm{\theta}, \bsm{\phi}} (\mbf{r}|\mbf{s})} \label{bayes-1} \\
	&= \frac{\int p(\mbf{x}_{1:K}, y) p_{\bsm{\theta}, \bsm{\phi}}(\mbf{r}|\mbf{x}_{1:K}, \mbf{s}) \mathrm{d} \mbf{x}_{1:K}}
	{\int p(\mbf{x}_{1:K}, y) p_{\bsm{\theta}, \bsm{\phi}}(\mbf{r}|\mbf{x}_{1:K}, \mbf{s}) \mathrm{d} \mbf{x}_{1:K} \mathrm{d} y}. \label{bayes-2}
\end{align}

\subsection{E2E Design Problem}
We are interested in finding the distributions $p_{\bsm{\theta}}(\mbf{z}_{1:K}|\mbf{x}_{1:K})$ and $p_{\bsm{\phi}}(\mbf{t}_{1:K}|\mbf{z}_{1:K}, \mbf{s})$ such that for each given state $\mbf{s}$, the received signal $\mbf{r}$ contains maximal information of the target $y$.
In other words, we aim to maximize the conditional mutual information $I(R;Y|S)$ w.r.t. $\bsm{\theta}$ and $\bsm{\phi}$, formulated as
\begin{align} 
	\text{(P1):}~\max_{\bsm{\theta}, \bsm{\phi}} \quad I(R;Y|S).
\end{align}
Note that
\begin{align}
	I(R;Y|S) &= H(Y|S) - H(Y|R,S) \\
	&= H(Y) - H(Y|R, S),
\end{align}
where the second equality holds due to the independence of $Y$ and $S$.
By ignoring the constant term $H(Y)$, the maximization of $I(R;Y|S)$ can be reformulated as
\begin{align} \label{expectations}
	\min_{\bsm{\theta}, \bsm{\phi}} \quad H(Y| R,S) =& \; \mathbb{E}_{p(\mbf{x}_{1:K}, y)} 
	\big[\mathbb{E}_{p(\mbf{r}| \mbf{t}_{1:K}, \mbf{s})}  \big[ \nonumber \\
	& \; \mathbb{E}_{p(\mbf{s})}
	\left[- \log p_{\bsm{\theta}, \bsm{\phi}} (y|\mbf{\mbf{r}}, \mbf{s})\right]
	\big]\big],
\end{align}
in which the objective $H(Y| R,S)$ characterizes the expected uncertainty of the inference result $Y$ for different realizations of $R$ and $S$. 
The expectations $\mathbb{E}_{p_{\bsm{\theta}}(\mbf{z}_{1:K}|\mbf{x}_{1:K})} \left[\cdot\right]$ and $\mathbb{E}_{p_{\bsm{\phi}}(\mbf{t}_{1:K}|\mbf{z}_{1:K}, \mbf{s})} \left[\cdot\right]$ are omitted in \eqref{expectations} by noting the deterministic forms of the feature encoder and precoder.
Recall \eqref{bayes-2}, the posterior $p_{\bsm{\theta}, \bsm{\phi}} (y|\mbf{r}, \mbf{s})$ in \eqref{expectations} is in general intractable due to the high-dimensional integrals with unknown distributions.
To tackle this issue, it is widely adopted to replace $p_{\bsm{\theta}, \bsm{\phi}} (y|\mbf{r}, \mbf{s})$ by a variational distribution with additional parameters to be learned \cite{VAE2014}.
The above formulation, built upon the explicit characterization of the state $S$, generalizes the work in \cite{jiawei2021task, jiawei2023multidevice, songjie2023digital, infomax2023sensors} where the physical-layer transmission is oversimplified as error-free bit pipes or AWGN channels. 
However, there are two main challenges that restrict the direct use of the above formulation to the E2E design of feature encoding and precoding:
\begin{itemize}
	\item Firstly, as indicated in \eqref{expectations}, evaluating $H(Y|R,S)$ requires to draw a sufficiently large number of samples simultaneously from the high-dimensional data distribution $p(\mbf{x}_{1:K}, y)$, the noise distribution $p(\mbf{r}|\mbf{t}_{1:K}, \mbf{s})$, and the channel state $p(\mbf{s})$.
	The simultaneous sampling over these distributions/datasets may incur a prohibitively large training overhead and unpredictable training complexity.
	\item Secondly,
	it is not clear how to design a network to achieve decent representation and generalization capabilities.
	The problem formulation \eqref{expectations} itself, unfortunately, does not provide any guidance regarding the selection of an appropriate network architecture.
	Although some heavily engineered networks (such as ResNet \cite{he2016deep} and ViT \cite{ViT2021ICLR}) have achieved empirical success for feature encoding in machine learning literature, the network architecture suitable for precoding in the considered task-oriented communication system is not well understood yet.
\end{itemize}

In this paper, we propose a decoupled pretraining framework that separately trains the feature encoder and the MIMO precoder prior to E2E learning.
With this decoupling, the feature encoding design does not involve the variation of the channel state.
On the other hand, the precoding design does not rely on the individual training samples in the task dataset.
We hence eliminate the need for simultaneous sampling over these distributions/datasets during the decoupled design phase.
We manage to align the pretraining objectives of each individual component with the E2E learning objective.
Since the decoupled design result can already achieve a decent performance owing to the aligned design objectives,  
leveraging it as the initialization for E2E learning effectively alleviate the huge training burden.
Regarding the second challenge, by incorporating prior knowledge of the feature distribution, we obtain a closed-form surrogate of the mutual information in (P1), known as coding rate reduction \cite{yu2020learning}.
This closed-form objective enables the development of a precoding optimization algorithm.
Iteratively unfolding the algorithm results in a network architecture tailored for this specific problem. 

\section{Decoupled Design of Feature Encoding, Precoding, and Classification} \label{decoupled-design}
In this section, we introduce a decoupled design framework for feature encoding and precoding based on the E2E formulation in (P1).
By introducing a Gaussian mixture (GM) prior on the learned features, this framework enables the derivation of a closed-form, Bayes-optimal classifier, eliminating the need for calculating high-dimensional integrals.

\begin{figure}
	[t]
	\centering
	\includegraphics[width=1\columnwidth]{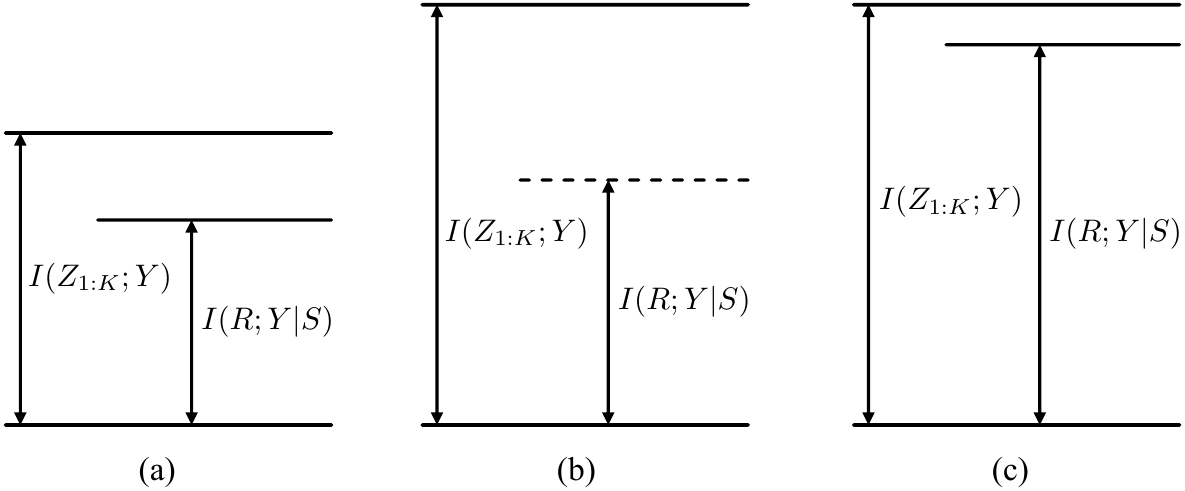}
	\caption{(a) Illustration of the data processing inequality in \eqref{data-processing}; 
		(b) The first step aims to maximize $I(Z_{1:K}; Y)$, while $I(R;Y|S)$ can either increase or decrease;
		(c) The second step aims to maximize $I(R;Y|S)$.}
	\label{data_processing}
\end{figure}

The idea stems from the data processing inequality \cite{cover1999elements}:
\begin{align}\label{data-processing}
	I(R;Y|S) 
	\leq I(Z_{1:K}; Y|S) 
	= I(Z_{1:K}; Y) ,
\end{align}
where the equality holds since the channel state $S$ is independent of the target $Y$ and the encoded features $Z_{1:K}$.
Given the ultimate goal of maximizing $I(R;Y|S)$, the inequality in \eqref{data-processing} implies the need for an even higher value of $I(Z_{1:K}; Y)$.
This motivates us to firstly optimize the upper bound $I(Z_{1:K}; Y)$, in which the optimization is only related to $p_{\bsm{\theta}}(\mbf{z}_{1:K}|\mbf{x}_{1:K})$, corresponding to the decoupled feature encoding problem.
However, optimizing the upper bound $I(Z_{1:K}; Y)$ alone does not ensure an improvement in the original objective $I(R;Y|S)$. 
Therefore, in the second step, we directly maximize $I(R;Y|S)$ w.r.t. $p_{\bsm{\phi}}(\mbf{t}_{1:K}|\mbf{z}_{1:K}, \mbf{s})$ with the feature encoder $p_{\bsm{\theta}}(\mbf{z}_{1:K}|\mbf{x}_{1:K})$ fixed, corresponding to the decoupled MIMO precoding problem. 
Fig. \ref{data_processing} summarizes the main idea of the decoupled two-step design framework.
Notably, as justified by \eqref{data-processing}, this framework establishes more aligned objectives of feature encoding and MIMO precoding compared to existing studies \cite{zhijin2021single, zhijin2021multiuser, zhang2023scan, wu2023deep}.
We elaborate the implementation details in the following subsections.

\subsection{Feature Encoding}
We now focus on the feature encoding problem: 
\begin{align} 
	\max_{\bsm{\theta}} \quad I(Z_{1:K}; Y).
\end{align}
By
\begin{align} \label{MI-decompose-1}
	I(Z_{1:K}; Y) = H(Y) - H(Y|Z_{1:K}),
\end{align}
one may consider to reformulate the above problem to
\begin{align} 
	\min_{\bsm{\theta}} \quad H(Y|Z_{1:K}) 
	= \mathbb{E}_{p(\mbf{x}_{1:K}, y)} \left[
	- \log p_{\bsm{\theta}} (y|\mbf{z}_{1:K}) \right] .
\end{align}
Similar to \eqref{bayes-2}, the posterior $p_{\bsm{\theta}} (y|\mbf{z}_{1:K})$ can be expressed as
\begin{align}
	p_{\bsm{\theta}} (y|\mbf{z}_{1:K}) = \frac{\int p(\mbf{x}_{1:K}, y) p_{\bsm{\theta}}(\mbf{z}_{1:K}|\mbf{x}_{1:K}) \mathrm{d} \mbf{x}_{1:K}}{\int p(\mbf{x}_{1:K}, y) p_{\bsm{\theta}}(\mbf{z}_{1:K}|\mbf{x}_{1:K}) \mathrm{d} \mbf{x}_{1:K} \mathrm{d} y},
\end{align}
which is in general intractable due to the high-dimensional integrals.
To overcome the difficulty arising in posterior computation, the variational method \cite{VAE2014} is often adopted to replace the true posterior $p_{\bsm{\theta}} (y|\mbf{z}_{1:K})$ by a variational approximation with some learnable parameters.
This approach justifies the minimization of the cross-entropy loss oftentimes seen in semantic communication literature \cite{zhijin2021single, zhijin2021multiuser, mxtao2023jsac, pingzhang2022jsac, jiawei2021task, jiawei2023multidevice, songjie2023digital, infomax2023sensors}.
However, in this approach, the precise geometric and statistical properties of the learned features $\mbf{z}_{1:K}$ are obscured.
It can hardly provide any useful model knowledge to guide the precoding design.
In what follows, we address such limitation by reformulating the objective towards learning statistically interpretable features from the data, so that the precoding design can benefit from the known statistics of the features. 

To begin with, we make the following key assumption on the prior $p(\mbf{z}_{1:K})$ and the likelihood $p(\mbf{z}_{1:K}|y)$, which is widely adopted in the literature \cite{GM2023Wen, cai2023multi}:
\begin{assump}
	Assume a finite number of classes, denoted by $J$, and let $\mathcal{J}\triangleq\{1, \dots, J\}$.
	The features follow a GM distribution, with each Gaussian component corresponding to a distinct class, i.e., 
	\begin{align} \label{z-GM}
		p(\mbf{z}_{1:K}) = \sum_{j \in \mathcal{J}}p_j p(\mbf{z}_{1:K}|y=j)
		= \sum_{j \in \mathcal{J}}p_j \mathcal{CN}(\mbf{0}, \mbf{\Sigma}_j), 
	\end{align}
	where $p_j$ is short for $p(y = j)$,
	and $\mbf{\Sigma}_j$ denotes the covariance matrix of $\mbf{z}_{1:K}$ in class $j$. 
\end{assump}
With the GM prior, we obtain a tight upper bound of $I(Z_{1:K}; Y)$ when the covariance matrix of $Z_{1:K}$ (denoted by $\mbf{\Sigma}$) and that in different classes (i.e., $\mbf{\Sigma}_j$) are non-degenerate: 
\begin{align}
	I(Z_{1:K}; Y) &= h(Z_{1:K}) - h(Z_{1:K}|Y) \label{MI-decompose-1b} \\
	&= h(Z_{1:K}) - \sum_{j \in \mathcal{J}} p_j h(Z_{1:K}|Y=j) \label{MI-decompose-2} \\
	&\leq \log \det \left(\pi e \mbf{\Sigma}\right) -  \sum_{j\in \mathcal{J}} p_j \log \det \left(\pi e \mbf{\Sigma}_j\right). \label{MI-decompose-3}
\end{align}
The inequality holds since $\log \det \left(\pi e \mbf{\Sigma}\right)$ tightly upper bounds $h(Z_{1:K})$, and $\log \det \left(\pi e \mbf{\Sigma}_j\right)$ exactly characterizes $h(Z_{1:K}|Y=j)$ by recalling the Gaussian assumption of $p(\mbf{z}_{1:K}|y=j)$.
However, the differential entropy is not well defined for degenerate $\mbf{\Sigma}$ or $\mbf{\Sigma}_j$. 
To handle the degenerate and non-degenerate cases both at once, we resort to the coding rate \cite{ma2007segmentation} that serves as an effective alternative to differential entropy.
Consequently, $I(Z_{1:K}; Y)$ can be approximated by the difference of coding rate terms, known as the coding rate reduction objective \cite{yu2020learning}, which is expressed as 
\begin{align}\label{mcr2-feature-encoding}
	\Delta \mathcal{R} (Z_{1:K}; Y) =& \log \det \left(\mbf{I} + \frac{D}{\epsilon^2}\mbf{\Sigma}\right) \nonumber \\
	&- \sum_{j \in \mathcal{J}} p_j \log \det \left(\mbf{I} + \frac{D}{\epsilon^2}\mbf{\Sigma}_j \right),
\end{align}
where $D \triangleq \sum_{k\in \mathcal{K}} D_k$ is the dimension of the concatenated feature vector $\mbf{z} \in \mathbb{C}^{D}$,
and $\epsilon$ is the lossy coding precision.\footnote{For interested readers,  please refer to \cite{ma2007segmentation, yu2020learning} for a rigorous explanation of $\Delta \mathcal{R} (Z_{1:K}; Y)$ from the lossy compression perspective.} 
Intuitively, the coding rate reduction objective amends the differential entropy terms in \eqref{MI-decompose-3} by adding a scaled identity matrix to the potentially degenerate covariance matrices, so as to avoid numerical issues for log-determinant computation.

\begin{remk}
	The above interpretation establishes the close connection among three information-theoretic measures, mutual information, cross-entropy, and coding rate reduction.
	Starting with $I(Z_{1:K}; Y)$, we arrive at the cross-entropy loss by the decomposition in \eqref{MI-decompose-1} with the variational approximation.
	We can also arrive at the coding rate reduction objective by the decomposition in \eqref{MI-decompose-1b} with the GM assumption.
	This justifies the remarkably good performance of the coding rate reduction as the loss function \cite{yu2020learning} and the network construction guideline \cite{chan2022redunet} for classification problems.
\end{remk}

Given a total of $M$ training data $\{\mbf{x}^{m}, y^{m}\}_{m=1}^M$, 
we obtain the corresponding feature samples $\mbf{Z} \triangleq \left[\mbf{z}^{1}, \dots, \mbf{z}^{M}\right]$ by feeding $\mbf{X} \triangleq \left[\mbf{x}^{1}, \dots, \mbf{x}^{M}\right]$ into the feature encoding network.
Then, we replace $\mbf{\Sigma}$ in \eqref{mcr2-feature-encoding} by its sample average $\frac{1}{M}\mbf{Z}\mbf{Z}^{\sf H}$. 
With the label information, we can also construct the feature samples of each class, denoted by $\mbf{Z}_j$ for class $j$. 
Likewise, each $\mbf{\Sigma}_j$ in \eqref{mcr2-feature-encoding} is replaced by $\frac{1}{M_j}\mbf{Z}_j\mbf{Z}_j^{\sf H}$, where $M_j$ denotes the number of samples in class $j$.
Based on the above, we have the following empirical estimate of the coding rate reduction objective:
\begin{align} \label{empirical-learning}
	\Delta \widetilde{\mathcal{R}} (Z_{1:K}; Y) &= \log \det \left(\mbf{I} + \frac{D}{M \epsilon^2}\mbf{Z}\mbf{Z}^{\sf H}\right) \nonumber\\ 
	&\!\!- \sum_{j \in \mathcal{J}} \frac{M_j}{M} \log \det \left(\mbf{I} + \frac{D}{M_j \epsilon^2}\mbf{Z}_j\mbf{Z}_j^{\sf H} \right).
\end{align}
The above empirical estimate serves as the learning objective of the feature encoding problem, known as the maximal coding rate reduction (MCR$^2$) criterion \cite{yu2020learning}:
\begin{subequations} \label{mcr2}
	\begin{align}
		\text{(P2):}~\max_{\mbf{Z}(\bsm{\theta})} \quad & \Delta \widetilde{\mathcal{R}}(Z_{1:K}; Y) \\ 
		\operatorname{ s.t. } \quad
		& \big\|\mbf{z}^{m} \big\|_2^2 = 1, ~~m = 1, \dots, M, \label{sphere}
	\end{align}
\end{subequations}
in which \eqref{sphere} normalizes the feature samples such that different representations can be compared fairly.
In practice, mini-batches of $\mbf{Z}$ are used in $\Delta \widetilde{\mathcal{R}} (Z_{1:K}; Y)$ to save memory and computation costs.

\begin{remk}
	Learning via the MCR$^2$ objective involves evaluating and differentiating a significant number of log-determinant terms that grows linearly with the number of classes.
	To reduce the training complexity, we refer interested readers to a recent work \cite{baek2022efficient} for an efficient variational formulation of the MCR$^2$ objective, which scales much more gracefully with the number of classes and the problem dimension.
\end{remk}

\begin{algorithm}[t]\label{alg1}
	\KwIn{Training dataset $\{\mbf{x}^{m}, y^{m}\}_{m=1}^M$, batch size $B_1$, initialized parameters $\bsm{\theta}$, coding precision $\epsilon$;}
	\KwOut{Optimized parameters $\bsm{\theta}$, feature covariance $\mbf{\Sigma}$ and that of different classes $\left\{\mbf{\Sigma}_j\right\}_{j\in \mathcal{J}}$;}
	\Repeat{Convergence of parameters $\bsm{\theta}$}{
		Randomly select a mini-batch $\{\mbf{x}^{m}, y^{m}\}_{m=1}^{B_1}$; \\ 
		\ForPar{$m=1,\dots,B_1$}{
			Feed forward $\mbf{z}^{m} = f_{\bsm{\theta}}(\mbf{x}^{m})$; \\
		}
		Compute the mini-batch version of the loss \eqref{empirical-learning}; \\
		Update parameters $\bsm{\theta}$ via backpropagation;}
	\ForPar{$m=1,\dots,M$}{
		Feed forward $\mbf{z}^{m} = f_{\bsm{\theta}}(\mbf{x}^{m})$; \\
	}
	Compute $\mbf{\Sigma} = \frac{1}{M}\mbf{Z}\mbf{Z}^{\sf H}$ and $\mbf{\Sigma}_j = \frac{1}{M_j}\mbf{Z}_j\mbf{Z}_j^{\sf H}$, $\forall j \in \mathcal{J}$.
	\caption{Training Feature Encoding Network}
\end{algorithm}

The training procedures for feature encoding are summarized in Algorithm \ref{alg1}. 
After training, we calculate the feature covariance and that of different classes by $\mbf{\Sigma} = \frac{1}{M}\mbf{Z}\mbf{Z}^{\sf H}$ and $\mbf{\Sigma}_j = \frac{1}{M_j}\mbf{Z}_j\mbf{Z}_j^{\sf H}$, respectively.
The calculated feature statistics serve as crucial prior knowledge for precoding design, as elaborated in the following subsection.

\begin{remk}
	Algorithm \ref{alg1} effectively exploits the correlations among different views during training.
	Instead of individually training the feature encoders at each device, in Algorithm \ref{alg1},
	we concatenate the feature outputs $\mbf{z}_k \in \mathbb{C}^{D_k}$ at each device into a single vector $\mbf{z} = \left[\mbf{z}_1^{\sf H}, \dots, \mbf{z}_K^{\sf H}\right]^{\sf H} \in \mathbb{C}^D$.
	Then, we employ the MCR$^2$ objective \eqref{empirical-learning} measured on $\mbf{z}$ as the loss function to train the feature encoders at different devices together.
	The MCR$^2$ objective \eqref{empirical-learning} implicitly characterizes the correlations among different $\mbf{z}_k$, as the term $\frac{1}{M} \mbf{Z}\mbf{Z}^{\sf H}$ therein represents the sample average of the covariance matrix of the concatenated features.
	By back-propagating the gradients computed from \eqref{empirical-learning}, the feature encoders at each device can effectively capture the correlations among different views.
\end{remk}

\subsection{MIMO Precoding}
Given the GM assumption \eqref{z-GM} on the learned features, the received signal $\mbf{r}$ (with the state $\mbf{s}$ given) is GM distributed as well, as indicated by \eqref{signal-model}.
The PDF is expressed as
\begin{align} \label{GM-received-feature}
	p(\mbf{r}|\mbf{s}) &= \sum_{j \in \mathcal{J}}p_j p(\mbf{r}|y=j, \mbf{s}) \\
	&= \sum_{j \in \mathcal{J}}p_j  \mathcal{CN}(\mbf{r}; \mbf{0},
	\mbf{H}\mbf{V} \mbf{\Sigma}_j \mbf{V}^{\sf H} \mbf{H}^{\sf H} + \sigma^2 \mbf{I}).
\end{align}
The GM distributed $p(\mbf{r}|\mbf{s})$ motivates us to apply the decomposition in \eqref{MI-decompose-1b} once again, this time on the precoding objective:
\begin{align}
	I(R;Y|S) &= h(R|S) - h(R|Y,S) \\
	&= \mathbb{E}_{p(\mbf{s})} \left[h(R|S=\mbf{s}) - h(R|Y,S=\mbf{s})\right].
\end{align}
Then, parallel to the previous subsection, we introduce the coding rate reduction measured on the received $R$ and label $Y$ conditioned on the state $S$, denoted by $\Delta \mathcal{R} (R; Y| S)$, as the surrogate of $I(R;Y|S)$.
Specifically, we replace $\mbf{\Sigma}$ and $\mbf{\Sigma}_j$ in \eqref{mcr2-feature-encoding} by the covariance matrix of $\mbf{r}$ and that of different classes.
By rearranging the terms, we can explicitly express this objective as 
\begin{align}\label{mcr2-for-precoding}
	\!\!\!\Delta \mathcal{R} (R; Y| S) &\!= \mathbb{E}_{p(\mbf{H}, \sigma)} \bigg[
	\log \det \left(\gamma\mbf{I} + \alpha \mbf{H}\mbf{V} (\bsm{\phi})  \mbf{\Sigma} \mbf{V}^{\sf H} (\bsm{\phi})  \mbf{H}^{\sf H} \right) \nonumber \\
	& \!\!\!\!\!\!\!\!\!\!\!\!\!\!\!\!\!\!\!\!\!\! - \sum_{j \in \mathcal{J}} p_j
	\log \det \left( \gamma\mbf{I} + \alpha \mbf{H}\mbf{V} (\bsm{\phi})  \mbf{\Sigma}_j \mbf{V}^{\sf H} (\bsm{\phi}) \mbf{H}^{\sf H}  \right) \bigg],
\end{align}
where
$\alpha \triangleq \frac{N_{\sf r}}{ \epsilon^2}$ and $\gamma \triangleq 1+\alpha \sigma^2$.
Note that $\Delta \mathcal{R} (R; Y| S)$ does not rely on the individual feature samples, but only the statistics of the encoded features, which is available prior to precoding design.
In this way, the precoding design is decoupled from the feature encoding problem, dedicated to address unfavorable wireless propagation conditions characterized by the state $\mbf{H}$ and $\sigma$.

Given a total of $N$ channel samples $\{\mbf{H}^{n}\}_{n=1}^N$ and a total of $E$ noise levels $\{\sigma^{e}\}_{e=1}^E$, the empirical estimate of \eqref{mcr2-for-precoding} is expressed as 
\begin{align}
	&\Delta \widetilde{\mathcal{R}} (R; Y| S) \nonumber\\
	=& \frac{1}{NE}  \sum_{n=1}^N \sum_{e=1}^E 
	\bigg[ \log \det \left(\gamma^{e} \mbf{I} + \alpha \mbf{H}^{n}\mbf{V}^{n,e} \mbf{\Sigma} \left(\mbf{V}^{n,e}\right)^{\sf H} \big(\mbf{H}^{n}\big)^{\sf H} \right) \nonumber \\
	&- \sum_{j \in \mathcal{J}} p_j
	\log \det \left( \gamma^{e}\mbf{I} + \alpha\mbf{H}^{n}\mbf{V}^{n,e}  \mbf{\Sigma}_j \left(\mbf{V}^{n,e}\right)^{\sf H} \big(\mbf{H}^{n}\big)^{\sf H}\right) \bigg], \label{mcr2_precoding_ep}
\end{align}
where $\mbf{V}^{n,e} = g_{\bsm{\phi}}(\mbf{H}^{n}, \sigma^{e})$ denotes the output of the precoding network, 
and $\gamma^e = 1 + \alpha (\sigma^e)^2$.
This empirical estimate serves as the learning objective of the precoding problem, formulated as
\begin{subequations}
	\begin{align}
		\text{(P3):}~\max_{\left\{\mbf{V}_k(\bsm{\phi}_k)\right\}_{k\in \mathcal{K}}} ~&  
		\Delta \widetilde{\mathcal{R}} (R; Y| S) \label{mcr2-obj} \\ 
		\operatorname{ s.t. } ~~~~~
		& \mathrm{tr} \big(\mbf{V}_k \mbf{\Sigma}^{(kk)} \mbf{V}_k^{\sf H}\big) \leq P_k, ~~ k\in \mathcal{K},
	\end{align}
\end{subequations}
where $\mbf{\Sigma}^{(kk)} \in \mathbb{C}^{D_k \times D_k}$ is the covariance matrix of $\mbf{z}_k$, and $P_k$ denotes the power budget at device $k$.

We summarize the training details of the precoding network in Algorithm \ref{alg2}.
The noise level is fixed in each mini-batch and randomly selected for different mini-batches.
The network architecture design will be detailed in Sections \ref{DU-BCA} and \ref{DU-BCA-MM}. 

\begin{algorithm}[t]\label{alg2}
	\KwIn{Channel dataset $\{\mbf{H}^{n}\}_{n=1}^N$, noise levels $\{\sigma^{e}\}_{e=1}^E$, batch size $B_2$, initialized parameters $\bsm{\phi}$, feature covariance $\mbf{\Sigma}$ and that of different classes $\left\{\mbf{\Sigma}_j\right\}_{j\in \mathcal{J}}$, coding precision $\epsilon$;}
	\KwOut{Optimized parameters $\bsm{\phi}$;}
	\Repeat{Convergence of parameters $\bsm{\phi}$}{
		Randomly select a mini-batch $\{\mbf{H}^{n}\}_{n=1}^{B_2}$; \\
		Select a noise level $\sigma^{e}$; \\
		\ForPar{$n=1,\dots,B_2$}{
			Feed forward $\mbf{V}^{n,e} = g_{\bsm{\phi}}(\mbf{H}^{n}, \sigma^{e})$; \\
		}
		Compute the mini-batch version of the loss \eqref{mcr2_precoding_ep}; \\
		Update parameters $\bsm{\phi}$ via backpropagation;}
	\caption{Training MIMO Precoding Network}
\end{algorithm}

\newcounter{TempEqCnt} 
\setcounter{TempEqCnt}{\value{equation}}
\setcounter{equation}{35}
\begin{figure*}[b]
	\hrulefill
	\begin{align} \label{ce_empirical}
		\widetilde{H}(Y| R,S) 
		&= \frac{1}{MNEF}\sum_{m=1}^{M} \sum_{n=1}^{N} \sum_{e=1}^{E} \sum_{f=1}^{F}
		\left[- \log p_{\bsm{\theta}, \bsm{\phi}} (y^m|\mbf{r}^{m,n,e,f}, \mbf{H}^n, \sigma^e)\right] \nonumber \\
		&= \frac{1}{MNEF}\sum_{m=1}^{M} \sum_{n=1}^{N} \sum_{e=1}^{E} \sum_{f=1}^{F} \left[- \log \frac{ p_{y^m} p_{\bsm{\theta}, \bsm{\phi}}(\mbf{r}^{m,n,e,f}|y^m, \mbf{H}^n, \sigma^e)}{\sum_{j} p_j p_{\bsm{\theta}, \bsm{\phi}}(\mbf{r}^{m,n,e,f}|j, \mbf{H}^n, \sigma^e) }\right] \nonumber \\
		&= \frac{1}{MNEF}\sum_{m=1}^{M} \sum_{n=1}^{N} \sum_{e=1}^{E} \sum_{f=1}^{F} \left[- \log \frac{ p_{y^m} \mathcal{CN}\left(\mbf{r}^{m,n,e,f}; \mbf{0},
			\mbf{H}^{n}\mbf{V}^{n,e} \mbf{\Sigma}_{y^m} \left(\mbf{V}^{n,e}\right)^{\sf H} \left(\mbf{H}^{n}\right)^{\sf H} + \left(\sigma^{e}\right)^2 \mbf{I}\right)}
		{\sum_{j} p_j \mathcal{CN}\left(\mbf{r}^{m,n,e,f}; \mbf{0},
			\mbf{H}^{n}\mbf{V}^{n,e} \mbf{\Sigma}_j \left(\mbf{V}^{n,e}\right)^{\sf H} \left(\mbf{H}^{n}\right)^{\sf H} + \left(\sigma^{e}\right)^2 \mbf{I}\right) }\right].
	\end{align}
\end{figure*}
\setcounter{equation}{\value{TempEqCnt}}

\begin{remk}
	In our previous work \cite{cai2023multi}, we formulated the same decoupled design problem, i.e., (P2) and (P3), in a more heuristic way.
	The coding rate reduction objective therein is interpreted as a measure on the separableness of different classes of intermediate features.
	From this point of view, (P2) is formulated to guarantee maximally separated $\mbf{z}$ for different classes.
	While the received signal $\mbf{r}$, gone through wireless transmission, may not maintain the same level of separability as $\mbf{z}$, hence leading to a deteriorated inference accuracy.
	To resolve this issue, (P3) is proposed to promote the separability of different classes of $\mbf{r}$ by optimizing $\mbf{V}$ to compensate for channel distortion.
	
	In contrast to \cite{cai2023multi}, this work provides an information-theoretic interpretation to the MCR$^2$ formulations in (P2) and (P3).
	In particular, we first present a unified E2E formulation targeted at conditional mutual information maximization.
	To overcome the difficulties in E2E learning, we provide a practical method to decouple the design of feature encoding and precoding from the original problem, justified by the data processing inequality.
	We exploit the close relation between coding rate and differential entropy, and then arrive at the MCR$^2$ formulations in (P2) and (P3).
\end{remk}

\subsection{Classification} \label{classification}
Owing to the GM assumption of $\mbf{z}$, the Bayes-optimal MAP classifier eliminates the need to calculate high-dimensional integrals as required in \eqref{bayes-2}.
Instead, it can be implemented much more easily as
\begin{align}
	\hat{y} &= \arg \max_{j \in \mathcal{J}} ~ p (y=j|\mbf{r}, \mbf{s}) \\
	&= \arg \max_{j \in \mathcal{J}} ~ p_j p (\mbf{r}|y=j, \mbf{s}) \label{MAP-1}\\
	&= \arg \max_{j \in \mathcal{J}} ~  p_j \mathcal{CN}(\mbf{r}; \mbf{0},
	\mbf{H}\mbf{V} \mbf{\Sigma}_j \mbf{V}^{\sf H} \mbf{H}^{\sf H} + \sigma^2 \mbf{I}), \label{MAP}
\end{align}
where \eqref{MAP-1} applies the Bayes' law.

\section{E2E Learning of Feature Encoding, Precoding, and Classification} \label{Sec-E2E-Learning}
After decoupled pretraining, we propose to fine-tune the feature encoding and MIMO precoding networks by E2E learning.
The E2E learning adopts the MAP classifier in Section \ref{classification} for computing the posterior probability of each class.
For each noise level $\sigma^{e}$, we generate $F$ independent noise realizations $\{\mbf{n}^{e,f}\}_{f=1}^{F}$ according to the distribution $\mathcal{CN}\big(\mbf{0}, (\sigma^{e})^2 \mbf{I}\big)$.
Then, with the training dataset $\{\mbf{x}^{m}, y^{m}\}_{m=1}^{M}$, the channel samples $\{\mbf{H}^{n}\}_{n=1}^{N}$, and the different noise levels $\{\sigma^{e}\}_{e=1}^{E}$, 
the empirical estimate of the E2E loss \eqref{expectations} is given in \eqref{ce_empirical} at the bottom of the previous page,
where $\mbf{r}^{m,n,e,f} = \mbf{H}^{n} \mbf{V}^{n,e} \mbf{z}^{m} + \mbf{n}^{e, f}$ with $\mbf{z}^{m} = f_{\bsm{\theta}}(\mbf{x}^{m})$ and $\mbf{V}^{n,e} = g_{\bsm{\phi}}(\mbf{H}^{n}, \sigma^{e})$.
Since the Gaussian PDF is differentiable, \eqref{ce_empirical} can be readily implemented in existing machine learning libraries to facilitate automatic gradient calculation.

Apparently, the nested sampling procedure suggested by \eqref{ce_empirical} entails a heavy training overhead.
Algorithm \ref{alg3} simplifies the sampling process as follows.
For each mini-batch with batch size $B_3$, we randomly select the data and channel samples $\{\mbf{x}^{i}, y^{i}\}_{i=1}^{B_3}$ and $\{\mbf{H}^{i}\}_{i=1}^{B_3}$.
Then, we select a noise level $\sigma^e$ and independently draw $B_3$ noise samples $\{\mbf{n}^{e, i}\}_{i=1}^{B_3}$ from the distribution $\mathcal{CN}\big(\mbf{0}, (\sigma^{e})^2\mbf{I}\big)$.
We feed forward $\mbf{x}^i$ and $\left\{\mbf{H}^i, \sigma^e\right\}$ into the feature encoder and MIMO precoder to generate $\mbf{z}^i$ and $\mbf{V}^{i,e}$ by $\mbf{z}^i = f_{\bsm{\theta}}\left(\mbf{x}^i\right)$ and $\mbf{V}^{i,e} = g_{\bsm{\phi}}(\mbf{H}^{i}, \sigma^{e})$, respectively.
Then, each feature sample $\mbf{z}^i$ is paired with $\mbf{H}^i$, $\mbf{V}^{i,e}$, and $\mbf{n}^{e, i}$ with the same index $i$ to generate the received signal $\mbf{r}^{i,i,e,i} = \mbf{H}^{i} \mbf{V}^{i,e} \mbf{z}^{i} + \mbf{n}^{e, i}$ for loss computation.

\begin{algorithm}[t]\label{alg3}
	\KwIn{Training dataset $\{\mbf{x}^{m}, y^{m}\}_{m=1}^M$, channel dataset $\{\mbf{H}^{n}\}_{n=1}^N$, noise levels $\{\sigma^{e}\}_{e=1}^E$, batch size $B_3$, $\bsm{\theta}$ obatined in Algorithm \ref{alg1}, $\bsm{\phi}$ obtained in Algorithm \ref{alg2}, feature covariance $\mbf{\Sigma}$ and that of different classes $\left\{\mbf{\Sigma}_j\right\}_{j\in \mathcal{J}}$;}
	\KwOut{Fine-tuned parameters $\bsm{\theta}$, $\bsm{\phi}$;}
	\Repeat{Convergence of parameters $\bsm{\theta}$, $\bsm{\phi}$}{
		Randomly select mini-batches of data and channel samples $\{\mbf{x}^{i}, y^{i}\}_{i=1}^{B_3}$ and $\{\mbf{H}^{i}\}_{i=1}^{B_3}$; \\
		Select a noise level $\sigma^{e}$ and independently generate the noise samples $\{\mbf{n}^{e, i}\}_{i=1}^{B_3}$ from $\mathcal{CN}\big(\mbf{0}, (\sigma^{e})^2\mbf{I}\big)$; \\
		\ForPar{$i=1,\dots,B_3$}{
			Feed forward $\mbf{z}^{i} = f_{\bsm{\theta}}(\mbf{x}^{i})$, $\mbf{V}^{i,e} = g_{\bsm{\phi}}(\mbf{H}^{i}, \sigma^{e})$; \\
			Compute $\mbf{r}^{i, i, e, i} = \mbf{H}^{i} \mbf{V}^{i,e} \mbf{z}^{i} + \mbf{n}^{e, i}$; \\
		}
		Compute the mini-batch version of the loss \eqref{ce_empirical}; \\
		Update parameters $\bsm{\theta}$, $\bsm{\phi}$ via backpropagation;
	}
	\caption{E2E Learning of Feature Encoding, MIMO Precoding, and Classification}
\end{algorithm}

\setcounter{TempEqCnt}{\value{equation}}
\setcounter{equation}{41}
\begin{figure*}[b]
	\hrulefill
	\begin{alignat}{2}
		&\mbf{b}_k &&\triangleq \mbf{D}_k^{-1} \mathrm{vec} \left(\mbf{H}^{\sf H}_k \mbf{U} \mbf{W}_0 \left(\big(\mbf{\Sigma}^{\frac{1}{2}}\big)^{(k)} \right)^{\sf H}
		- \mbf{H}^{\sf H}_k \mbf{U} \mbf{W}_0 \mbf{U}^{\sf H}\sum_{q \neq k} \mbf{H}_q \mbf{V}_q \mbf{\Sigma}^{(qk)} 
		- \alpha \sum_{j\in\mathcal{J}} p_j \mbf{H}^{\sf H}_k \mbf{W}_j \sum_{q \neq k} \mbf{H}_q \mbf{V}_q \mbf{\Sigma}_j^{(qk)}\right), \label{T_k}\\
		&\mbf{N}_k &&\triangleq \mbf{D}_k^{-1}
		\left(  \big(\mbf{\Sigma}^{(kk)}\big)^{\sf T} \otimes \big(\mbf{H}_k^{\sf H} \mbf{U} \mbf{W}_0 \mbf{U}^{\sf H} \mbf{H}_k \big)
		+ \alpha \sum_{j\in\mathcal{J}} p_j \big(\mbf{\Sigma}_j^{(kk)}\big)^{\sf T} \otimes \big( \mbf{H}_k^{\sf H}\mbf{W}_j \mbf{H}_k \big) \right)
		\mbf{D}_k^{-1}. \label{M_k}
	\end{alignat}
\end{figure*}
\setcounter{equation}{\value{TempEqCnt}}
\section{Vanilla DU-BCA Precoder: Algorithm Foundation and Deep Unfolding} \label{DU-BCA}
The authors in \cite{cai2023multi} aim to solve (P3) as an optimization problem for each given $\mbf{H}$ and $\sigma$: 
\setcounter{equation}{36}
\begin{subequations} \label{P4}
	\begin{align}
		 \text{(P4):}~\max_{\left\{\mbf{V}_k\right\}_{k\in \mathcal{K}}} ~&   \log \det \left(\mbf{F}_0\right) - \sum_{j \in \mathcal{J}} p_j \log \det \left(\mbf{F}_j\right)
		 \\ 
		\operatorname{ s.t. } ~~~~~
		& \mathrm{tr} \big(\mbf{V}_k \mbf{\Sigma}^{(kk)} \mbf{V}_k^{\sf H}\big) \leq P_k, ~~ k\in \mathcal{K},
	\end{align}
\end{subequations}
where $\mbf{F}_0 \triangleq \gamma\mbf{I} + \alpha \mbf{H}\mbf{V} \mbf{\Sigma} \mbf{V}^{\sf H} \mbf{H}^{\sf H}$ and
$\mbf{F}_j \triangleq  \gamma\mbf{I} + \alpha \mbf{H}\mbf{V} \mbf{\Sigma}_j \mbf{V}^{\sf H} \mbf{H}^{\sf H}$.
In this section, we first outline the BCA solution algorithm proposed in \cite{cai2023multi}.
Building upon this, we propose a deep unfolding network for precoding design, which primarily follows the methodologies discussed in \cite{hu2020iterative, hu2021joint, liu2021deep}.
We then identify the inherent limitations of this simplistic approach, motivating the enhancements we are going to propose in the next section.

\subsection{BCA Algorithm}\label{Sec-BCA-Algorithm}
Let $\mathcal{V}_k \triangleq \big\{\mbf{V}_k \big| \mathrm{tr} \big(\mbf{V}_k \mbf{\Sigma}^{(kk)} \mbf{V}_k^{\sf H}\big) \leq P_k\big\}$ denote the feasible set of $\mbf{V}_k$.
By introducing the auxiliary variables $\left\{\mbf{W}_j \succ \mbf{0}\right\}_{j\in \mathcal{J}_0}$ and $\mbf{U}$ with $\mathcal{J}_0 \triangleq \left\{0\right\} \cup \mathcal{J}$, we can reformulate (P4) to the following problem \cite{cai2023multi}:
\begin{align}
	\text{(P5):}~\max_{\{\mbf{W}_j \succ \mbf{0}\}_{j\in \mathcal{J}_0}, \mbf{U}, \atop \{\mbf{V}_k \in \mathcal{V}_k\}_{k\in \mathcal{K}} }  & \log \det \left(\mbf{W}_0\right) - \mathrm{tr} \left(\mbf{W}_0 \mbf{E}_0\right)  \nonumber \\
	&\!\!\!\! \!\!\!\! \!\!\!\! \!\! \!\!\!\! \!\! \!\!\! \!\! \! + \sum_{j\in \mathcal{J}} p_j \left\{ \log \det \left(\mbf{W}_j\right) - \mathrm{tr} \left(\mbf{W}_j \mbf{F}_j\right) \right\} ,
\end{align}
where $\mbf{E}_0 \triangleq \big(\mbf{I} - \mbf{U}^{\sf H} \mbf{H} \mbf{V} \mbf{\Sigma}^{\frac{1}{2}} \big) \big(\mbf{I} - \mbf{U}^{\sf H} \mbf{H} \mbf{V} \mbf{\Sigma}^{\frac{1}{2}}\big)^{\sf H} + \frac{\gamma}{\alpha} \mbf{U}^{\sf H} \mbf{U}$.
It can be verified that (P4) and (P5) share the same optimal solution $\{\mbf{V}_k\}_{k\in \mathcal{K}}$  \cite{cai2023multi}.
The BCA algorithm is employed to solve (P5) by updating one block of variables at a time with the other blocks fixed.
It can be shown that (P5) is convex w.r.t. $\mbf{U}$ and $\{\mbf{W}_j\}_{j \in \mathcal{J}_0}$ individually, which results in the following two update steps in the BCA algorithm:
\begin{mdframed}[hidealllines=true, backgroundcolor=black!20]
	\vspace{-.6em}
	\begin{alignat}{3}
		&\text{($\mbf{U}$-step):} &&\mbf{U}  &&= \alpha \mbf{F}_0^{-1} \mbf{H} \mbf{V} \mbf{\Sigma}^{\frac{1}{2}},~~~~ \label{U-update}\\
		&{\text{($\mbf{W}$-step):}}~~~~&&\mbf{W}_j &&= \begin{cases}
			\mbf{E}_0^{-1}, &j = 0,  \\
			\mbf{F}_j^{-1}, &j \in \mathcal{J}.
		\end{cases} \label{W-update}
	\end{alignat}
\end{mdframed}

Due to the coupling of different $\mbf{V}_k$'s in (P5), it is required to sequentially optimize each precoder with the others fixed.
The sub-problem for optimizing $\mbf{V}_k$ is a convex quadratically constrained quadratic program (QCQP).
By letting $\mbf{v}_k \triangleq \mbf{D}_k \mathrm{vec} \left(\mbf{V}_k\right)$ with $\mbf{D}_k \triangleq \big((\mbf{\Sigma}^{(kk)})^{\sf T}\otimes \mbf{I} \big)^{\frac{1}{2}}$, the convex QCQP is expressed in its standard form as
\begin{subequations}
	\begin{align}
		\text{(P6):}~~\min_{\mbf{v}_k} ~~&  
		-2 \mathrm{Re} \left\{ \mbf{b}_k^{\sf H} \mbf{v}_k \right\}
		+ \mbf{v}_k^{\sf H}  \mbf{N}_k \mbf{v}_k \\ 
		\operatorname{ s.t. } ~~
		& \mbf{v}_k^{\sf H} \mbf{v}_k\leq P_k, \label{lagrangian-multiplier}
	\end{align}
\end{subequations}
where $\mbf{b}_k$ and $\mbf{N}_k$ are respectively given in \eqref{T_k} and \eqref{M_k} at the bottom of this page.
In \eqref{T_k}, $\big(\mbf{\Sigma}^{\frac{1}{2}}\big)^{(k)}$ denotes row $m_k$ to row $n_k$ of $\mbf{\Sigma}^{\frac{1}{2}}$, where $m_k \triangleq \sum_{i=0}^{k-1}D_i +1$ and $n_k \triangleq \sum_{i=1}^k D_i$, with $D_0 \triangleq 0$;
$\mbf{\Sigma}_{j}^{(qk)}$ is formed by row $m_q$ to row $n_q$ and column $m_k$ to column $n_k$ of $\mbf{\Sigma}_j$.
Solving (P6) for each device sequentially yields the following update rule: 
\setcounter{equation}{43}
\begin{mdframed}[hidealllines=true, backgroundcolor=black!20]
	\vspace{-.6em}
	\begin{subequations}\label{Vv-update}
		\begin{align}
			\!\!\!\!\!{\text{($\mbf{V}$-step):}}~~~\; &\textbf{for}~ k = 1, \dots, K \textbf{~do} \nonumber \\
			&~~~\mbf{v}_k \; = \left(\mbf{N}_k + \lambda_k \mbf{I}\right)^{-1} \mbf{b}_k; \label{v-update} \\
			&~~~\mbf{V}_k = \mathrm{vec}^{-1} \left( \mbf{D}_k^{-1}  \mbf{v}_k \right). \label{V-update} \\
			&\textbf{end} \nonumber 
		\end{align}
	\end{subequations}
\end{mdframed}
In \eqref{v-update}, $\lambda_k$ is the Lagrange multiplier associated with \eqref{lagrangian-multiplier}, which can be calculated via a bisection search.
In \eqref{V-update}, $\mathrm{vec}^{-1} (\cdot)$ denotes the inverse operation of $\mathrm{vec} (\cdot)$, which de-vectorizes the argument from a vector to a matrix.

To summarize, starting with an initial guess of $\mbf{V}$, the $\mbf{U}\text{-}$, $\mbf{W}$-, and $\mbf{V}$-steps in \eqref{U-update}, \eqref{W-update}, and \eqref{Vv-update} are iteratively executed till convergence.

\subsection{Vanilla DU-BCA Precoder} \label{sec-vanilla-precoder}
We unfold the BCA algorithm into a layer-wise structure with some learnable parameters introduced, which we refer to as the vanilla DU-BCA precoder.
From Section \ref{Sec-BCA-Algorithm}, we see that the BCA algorithm involves frequent calculation of matrix inverses in all the three update steps.
The matrix inverse operation imposes a computational complexity that scales cubically with the matrix dimension.
It is also noteworthy that the $\mbf{V}$-step necessitates the iterative adjustment of the Lagrange multipliers, an operation relying on the eigendecomposition with a cubic complexity.
Consequently, it is natural to replace these computational extensive operations by some learnable lightweight structures. 

Inspired by \cite{hu2020iterative}, we adopt the following structure with learnable parameters $\bsm{\Xi}_1, \bsm{\Xi}_2, \bsm{\Xi}_3 \in \mathbb{C}^{n\times n}$ to approximate the inversion of a given matrix $\mbf{A} \in \mathbb{C}^{n\times n}$:
\begin{align} \label{matrix-inverse-approx}
	\mbf{A}^{-1} \approx \mbf{A}^\ddagger \bsm{\Xi}_1 + \mbf{A} \bsm{\Xi}_2 + \bsm{\Xi}_3,
\end{align}
where $\mbf{A}^\ddagger \triangleq \diag \big\{\frac{1}{a_{11}}, \frac{1}{a_{22}}, \dots, \frac{1}{a_{nn}}\big\}$ takes the reciprocal of the diagonal elements in $\mbf{A}$ and sets the off-diagonal elements as zero.
Two justifications are given for the approximation in \eqref{matrix-inverse-approx}.
Firstly, $\mbf{A}^\ddagger$ itself is a good estimation of $\mbf{A}^{-1}$ when $\mbf{A}$ is diagonally dominant, i.e., the diagonal elements of $\mbf{A}$ are much larger than the off-diagonal elements.
This leads to the term $\mbf{A}^\ddagger \bsm{\Xi}_1$ in \eqref{matrix-inverse-approx}.
Secondly, $\mbf{A}\bsm{\Xi}_2 + \bsm{\Xi}_3$ resembles the first-order Taylor expansion of $\mbf{A}^{-1}$ at $\mbf{A}_0$: $\mbf{A}^{-1} \approx 2 \mbf{A}_0^{-1} - \mbf{A}_0^{-1}\mbf{A}\mbf{A}_0^{-1}$.
We apply the approximation in \eqref{matrix-inverse-approx} to learn the updates of $\mbf{U}$ and $\left\{\mbf{W}_j\right\}_{j \in \mathcal{J}_0}$ at the $\ell$-th layer as
\begin{mdframed}[hidealllines=true, backgroundcolor=black!20]
	\vspace{-.6em}
	\begin{alignat}{2}
		&\mbf{U}^\ell  &&= \alpha \left(\left(\mbf{F}_0^\ell\right)^\ddagger \bsm{\Theta}_{1}^{\ell} + \mbf{F}_0^\ell \bsm{\Theta}_{2}^{\ell} + \bsm{\Theta}_{3}^{\ell} \right) \mbf{H} \mbf{V}^\ell \mbf{\Sigma}^{\frac{1}{2}}, \label{U-learnable-update}\\
		&\mbf{W}_j^\ell &&= \begin{cases}
			\left(\mbf{E}_0^\ell\right)^\ddagger \bsm{\Phi}_{1}^{\ell} +\mbf{E}_0^\ell \bsm{\Phi}_{2}^{\ell} + \bsm{\Phi}_{3}^{\ell}, &j = 0,  \\
			\left(\mbf{F}_j^\ell\right)^\ddagger \bsm{\Psi}_{1}^{\ell}  +\mbf{F}_j^\ell \bsm{\Psi}_{2}^{\ell} + \bsm{\Psi}_{3}^{\ell}, &j \in \mathcal{J},
		\end{cases} \label{W-learnable-update}
	\end{alignat}
\end{mdframed}
where $\bsm{\Theta}_1^\ell, \bsm{\Theta}_2^\ell, \bsm{\Theta}_3^\ell \in \mathbb{C}^{N_{\sf r} \times N_{\sf r}}$ denote the learnable parameters introduced to approximate $(\mbf{F}_0^\ell)^{-1}$;
$\bsm{\Phi}_1^\ell, \bsm{\Phi}_2^\ell, \bsm{\Phi}_3^\ell \in \mathbb{C}^{D \times D}$ denote the learnable parameters introduced to approximate $(\mbf{E}_0^\ell)^{-1}$;
and $\bsm{\Psi}_1^\ell, \bsm{\Psi}_2^\ell, \bsm{\Psi}_3^\ell \in \mathbb{C}^{N_{\sf r} \times N_{\sf r}}$ denote are the learnable parameters introduced to approximate $(\mbf{F}_j^\ell)^{-1}$.
Note that we reuse the matrix inversion approximator for different classes of $\mbf{F}_j^\ell$ to reduce the network size (number of learnable parameters).
In the sequel, we use $\bsm{\Theta}^\ell \triangleq \{\bsm{\Theta}_1^\ell, \bsm{\Theta}_2^\ell, \bsm{\Theta}_3^\ell\}$,
$\bsm{\Phi}^\ell \triangleq \{\bsm{\Phi}_1^\ell, \bsm{\Phi}_2^\ell, \bsm{\Phi}_3^\ell\}$, and
$\bsm{\Psi}^\ell \triangleq \{\bsm{\Psi}_1^\ell, \bsm{\Psi}_2^\ell, \bsm{\Psi}_3^\ell\}$ for abbreviation.

\begin{figure}
	[t]
	\centering
	\includegraphics[width=1\columnwidth]{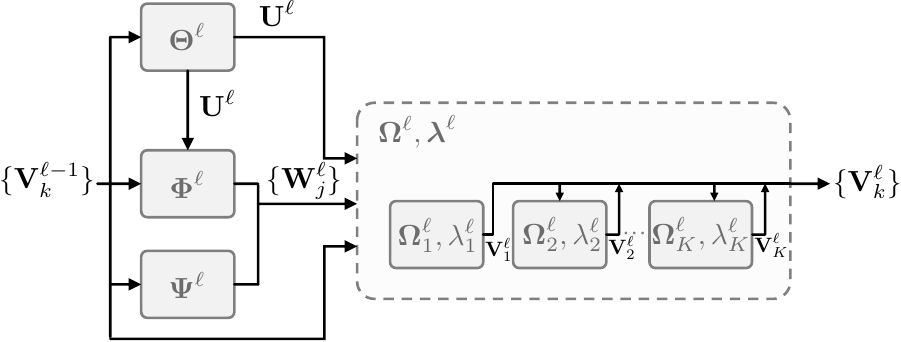}
	\caption{The $\ell$-th layer of the vanilla DU-BCA precoder.}
\label{vanilla}
\end{figure}

For the update of $\left\{\mbf{V}_k\right\}_{k \in \mathcal{K}}$, apart from the matrix inversion, the calculation of the Lagrange multipliers $\{\lambda_k\}_{k \in \mathcal{K}}$ entails a time-consuming eigendecomposition, followed by a bisection search that are complicated to represent as standard NN structures.
A potential solution to address this issue is to also set $\{\lambda_k\}_{k \in \mathcal{K}}$ as learnable parameters \cite{hu2021joint, liu2021deep}.
The matrix inversion in \eqref{v-update} can be parameterized as
\begin{align}\label{over-para}
	\left(\mbf{N}_k + \lambda_k \mbf{I}\right)^{-1} &\approx \left(\mbf{N}_k + \lambda_k \mbf{I}\right)^{\ddagger} \bsm{\Xi}_1 + \left(\mbf{N}_k + \lambda_k \mbf{I}\right) \bsm{\Xi}_2 + \bsm{\Xi}_3 \nonumber \\
	&= \left(\mbf{N}_k + \lambda_k \mbf{I}\right)^{\ddagger} \bsm{\Xi}_1 + \mbf{N}_k \bsm{\Xi}_2  + \left(\lambda_k \bsm{\Xi}_2 + \bsm{\Xi}_3\right). 
\end{align}
As shown in \eqref{over-para}, $\lambda_k \bsm{\Xi}_2$ in $\left(\mbf{N}_k + \lambda_k \mbf{I}\right) \bsm{\Xi}_2$ can be merged to the third term.
Therefore, we turn to parameterize $\left(\mbf{N}_k + \lambda_k \mbf{I}\right)^{-1}$ more concisely as 
\begin{align}
	\left(\mbf{N}_k + \lambda_k \mbf{I}\right)^{-1} \approx \left(\mbf{N}_k + \lambda_k \mbf{I}\right)^{\ddagger} \bsm{\Xi}_1 + \mbf{N}_k \bsm{\Xi}_2 + \bsm{\Xi}_3.
\end{align}
Based on the above parameterization, the update of $\{\mbf{V}_k\}_{k\in\mathcal{K}}$ at the $\ell$-th layer is constructed as
\begin{mdframed}[hidealllines=true, backgroundcolor=black!20]
		\vspace{-.6em}
	\begin{subequations}\label{Vv-learnable-update}
		\begin{align}
			&\!\!\!\!\!\!\textbf{for}~ k = 1, \dots, K \textbf{~do} \nonumber \\
			&~\mbf{v}_k^\ell \; = \big(\!\left(\mbf{N}_k^\ell \!+\! \lambda_k^\ell \mbf{I}\right)^{\ddagger} \bsm{\Omega}_{k, 1}^\ell \!+\! \mbf{N}_k^\ell \bsm{\Omega}_{k,2}^\ell \!+\! \bsm{\Omega}_{k,3}^\ell \big) \mbf{b}_k^\ell; \label{v-learnable-update} \\  
			&~\mbf{V}_k^\ell  = \mathrm{Proj}_{\mathcal{V}_k} \left\{\mathrm{vec}^{-1} \left( \mbf{D}_k^{-1}  \mbf{v}_k^\ell \right)\right\}; \label{V-learnable-update} \\
			&\!\!\!\!\!\!\textbf{end} \nonumber 
		\end{align}
	\end{subequations}
\end{mdframed}
In \eqref{v-learnable-update}, $\bsm{\Omega}_{k, 1}^\ell, \bsm{\Omega}_{k, 2}^\ell, \bsm{\Omega}_{k, 3}^\ell \in \mathbb{C}^{D_k N_{{\sf t}, k} \times D_k N_{{\sf t}, k}}$ and $\lambda_k^\ell \in \mathbb{C}$ are the introduced learnable parameters. 
For notation simplicity, define $\bsm{\Omega}^\ell_k \triangleq \{\bsm{\Omega}_{k,1}^\ell, \bsm{\Omega}_{k,2}^\ell, \bsm{\Omega}_{k,3}^\ell\}$,
$\bsm{\Omega}^\ell \triangleq \{\bsm{\Omega}^\ell_k\}_{k \in \mathcal{K}}$, and
$\bsm{\lambda}^\ell \triangleq \{\lambda_k^\ell\}_{k\in\mathcal{K}}$.
Since \eqref{v-learnable-update} is an approximation of the update in \eqref{v-update},
the resulting $\mbf{V}_k^\ell$ may not always adhere to the power constraint.
To tackle this issue, we append a projection operator in \eqref{V-learnable-update}, defined as
\begin{align} \label{projection}
	\mathrm{Proj}_{\mathcal{V}_k} \left\{\mbf{V}_k\right\} \triangleq 
	\begin{cases}
		\mbf{V}_k,  & \mbf{V}_k \in \mathcal{V}_k, \\
		\sqrt{\frac{P_k}{\mathrm{tr} \left(\mbf{V}_k \mbf{\Sigma}^{(kk)} \mbf{V}_k^{\sf H}\right)}} \mbf{V}_k, & \text{otherwise}.
	\end{cases}
\end{align}
In \eqref{V-learnable-update}, $\mbf{D}_k^{-1}$ is irrelevant to the iterations/layers by recalling the definition $\mbf{D}_k \triangleq \big((\mbf{\Sigma}^{(kk)})^{\sf T}\otimes \mbf{I} \big)^{\frac{1}{2}}$, and hence can be calculated offline.
Fig. \ref{vanilla} depicts one layer of the vanilla DU-BCA precoder, which consists of the components defined in \eqref{U-learnable-update}, \eqref{W-learnable-update}, and \eqref{Vv-learnable-update}.

There are three main issues that potentially limit the performance of the vanilla DU-BCA precoder proposed in this section. 
Firstly, the matrix inversion approximator in \eqref{matrix-inverse-approx} may be inaccurate for general matrices with non-negligible off-diagonal elements.
We have empirically observed that the matrices for inverse calculation in the $\mbf{U}$- and $\mbf{W}$-steps are diagonally dominant, while these in the $\mbf{V}$-step are not.
Secondly, treating the Lagrange multipliers directly as learnable parameters implies their constant values for different channel realizations, which loses the freedom for adaptive adjustment as in the optimization algorithm counterpart.
Thirdly, the role of Lagrange multipliers is to penalize the optimization objective for the satisfaction of the constraints. 
However, as the optimization objective itself usually serves as the loss function to train the deep unfolding network, updating the learnable Lagrange multipliers via backpropagation directs them to improve the objective value of (P4) following gradient ascent.
This contradicts the intended purpose for penalization. 

\section{Enhanced DU-BCA-MM Precoder: Algorithm Development and Deep Unfolding}\label{DU-BCA-MM}
In this section, we develop an alternative update rule for the $\mbf{V}$-step in \eqref{Vv-update}, which circumvents the non-diagonal matrix inversion and the Lagrangian tuning process. 
Based on the developed algorithm, we propose a new deep unfolding network to improve the performance of the vanilla DU-BCA precoder.

\subsection{BCA-MM Algorithm}\label{Sec-BCA-MM} 
This subsection introduces a novel approach to solve (P6).
We begin with the following useful result.
\begin{prop}[\cite{palomar2018icassp}]
	Let $\mbf{L}, \mbf{M} \in \mathbb{H}^n$ such that $\mbf{M} \succeq \mbf{L}$.
	The function $\mbf{v}^{\sf H} \mbf{L} \mbf{v}$ with $\mbf{v} \in \mathbb{C}^n$ is majorized at any point $\underline{\mbf{v}} \in \mathbb{C}^n$ by
	\begin{align} \label{majorization}
		\mbf{v}^{\sf H} \mbf{L} \mbf{v} 
		\leq \mbf{v}^{\sf H} \mbf{M} \mbf{v}
		+ 2 \mathrm{Re} \left\{\mbf{v}^{\sf H} \left(\mbf{L} - \mbf{M}\right) \underline{\mbf{v}}\right\}
		+ \underline{\mbf{v}}^{\sf H} \left(\mbf{M} - \mbf{L}\right) \underline{\mbf{v}}.
	\end{align}
\end{prop}
By choosing $\mbf{L} = \mbf{N}_k$ and $\mbf{M} = \eta_k \mbf{I}$ with $\eta_k \geq \lambda_{\max} \left(\mbf{N}_k\right)$, we construct an upper bound of the objective in (P6) at any given point $\underline{\mbf{v}}_k$ as
\begin{align}\label{upper-bound}
	u_k (\mbf{v}_k | \underline{\mbf{v}}_k ) =&\;
	\eta_k \mbf{v}_k^{\sf H} \mbf{v}_k  - 2 \mathrm{Re} \big\{\left(\mbf{b}_k - \left(\mbf{N}_k - \eta_k \mbf{I}\right) \underline{\mbf{v}}_k \right)^{\sf H} \mbf{v}_k\big\} \nonumber \\
	& + \underline{\mbf{v}}_k^{\sf H} \left(\eta_k \mbf{I}-\mbf{N}_k\right) \underline{\mbf{v}}_k.
\end{align}
The Perron-Frobenius Theorem \cite[Corollary A4]{garren1968bounds} provides a computationally efficient choice for $\eta_k$ by setting $\eta_k$ as the maximum absolute row sum of $\mbf{N}_k$.
That is, $\eta_k =  \max_i \sum_j \left|n_{k, ij}\right|$, where $n_{k,ij}$ denotes the $(i,j)$-th element in $\mbf{N}_k$.

We propose an MM algorithm to iteratively minimize the upper bound in \eqref{upper-bound}, where $\underline{\mbf{v}}_k$ in each iteration is set to be the solution obtained from the last iteration.
This iterative process converges to the optimal solution to (P6) owing to its convexity.
By ignoring the terms irrelevant to optimization, we write the problem to be solved in each MM iteration as
\begin{subequations}
	\begin{align}
		\text{(P7):}~~\min_{\mbf{v}_k} ~~&  
		\eta_k \left\|\mbf{v}_k - \mbf{q}_k\right\|_2^2 \\ 
		\operatorname{ s.t. } ~~
		& \mbf{v}_k^{\sf H} \mbf{v}_k\leq P_k,
	\end{align}
\end{subequations}
where
\begin{align} \label{q-update}
	\mbf{q}_k \triangleq \frac{1}{\eta_k}\left(\mbf{b}_k - \left(\mbf{N}_k -\eta_k\mbf{I}\right)\underline{\mbf{v}}_k\right).
\end{align}
Fortunately, the closed-form solution to (P7) can be obtained without introducing Lagrange multipliers.
It is given in a matrix-inversion-free form as
\begin{align}
	\mbf{v}_k = \mbf{q}_k \min \left\{ \frac{\sqrt{P_k}}{\left\| \mbf{q}_k \right\|_2} , 1 \right\}.
\end{align}
For notation simplicity, we assume that the same number of iterations, denoted by $I$, is required to reach convergence for all $k$ in the MM algorithm.
The MM-based implementation of the $\mbf{V}$-step is given by
\begin{mdframed}[hidealllines=true, backgroundcolor=black!20]
	\vspace{-.6em}
	\begin{subequations}\label{Vv-MM-update}
		\begin{align}
			&\textbf{for}~ k = 1, \dots, K \textbf{~do} \nonumber \\ 
			&~~~\textbf{for}~ i =1, \dots, I \textbf{~do} \nonumber \\
			&~~~~~~\mbf{q}_k^i \; = \frac{1}{\eta_k}\left(\mbf{b}_k - \left(\mbf{N}_k -\eta_k\mbf{I}\right)\mbf{v}_k^{i-1}\right); \\
			&~~~~~~\mbf{v}_k^i \; = \mbf{q}_k^i \min \left\{ \frac{\sqrt{P_k}}{\left\| \mbf{q}_k^i \right\|_2} , 1 \right\}; \label{v-MM-update} \\
			&~~~\textbf{end} \nonumber \\
			&~~~\mbf{V}_k = \mathrm{vec}^{-1} \left( \mbf{D}_k^{-1}  \mbf{v}_k^I \right); \label{V-MM-update} \\
			&\textbf{end} \nonumber 
		\end{align}
	\end{subequations}
\end{mdframed}
In summary, the BCA-MM algorithm integrates the $\mbf{U}$- and $\mbf{W}$-steps in \eqref{U-update} and \eqref{W-update}, together with the modified $\mbf{V}$-step in \eqref{Vv-MM-update}, to facilitate iterative updates. 

\subsection{Enhanced DU-BCA-MM Precoder}
To address the limitations of the vanilla DU-BCA precoder,
we propose a new deep unfolding architecture for updating $\{\mbf{V}_k\}_{k\in\mathcal{K}}$ based on the MM algorithm.
The modified $\mbf{V}$-step in \eqref{Vv-MM-update} involves a nested iteration, which may lead to a relatively slow convergence speed.
In the context of deep unfolding, this implies the need of a large number of sub-layers to achieve satisfactory performance, incurring a scalability issue.

\begin{figure}
	[t]
	\centering
	\includegraphics[width=1\columnwidth]{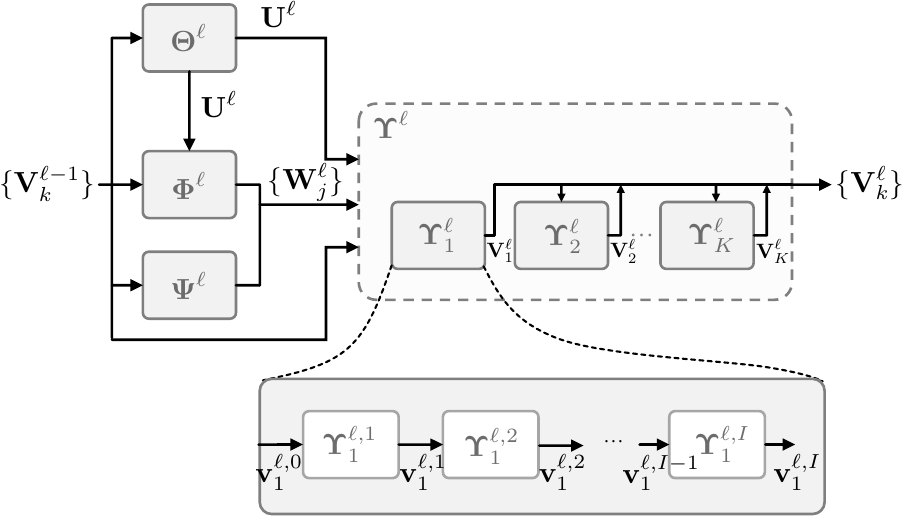}
	\caption{The $\ell$-th layer of the enhanced DU-BCA-MM precoder.
	}
	\label{proposed-DU-BCA-MM}
\end{figure}

Our design aims to introduce some learnable parameters to accelerate the convergence of the MM algorithm.
Notice that the choice of $\mbf{M}$ in \eqref{majorization} significantly influences the convergence speed as it controls the tightness of the bound. 
Even if we choose $\mbf{M} = \lambda_{\max} \left(\mbf{N}_k\right) \mbf{I}$ regardless of the associated complexity of computing $\lambda_{\max} \left(\mbf{N}_k\right)$, the bound can still be loose since the other eigenvalues of $\mbf{N}_k$ may be much smaller than $\lambda_{\max} \left(\mbf{N}_k\right)$.
On the other hand, in order to obtain the closed-form solution in each MM iteration, it is necessary to maintain $\mbf{M}$ as a scaled identity matrix, i.e., $\eta_k \mbf{I}$.
Given the inherent limitation in constructing upper bounds, we focus on learning the $\mbf{M}$ matrix that optimally conforms to the update rule in \eqref{q-update}.
Specifically, we replace the identity matrix in \eqref{q-update} by a learnable matrix $\bsm{\Upsilon}_k\in \mathbb{C}^{D_k N_{{\sf t}, k} \times D_k N_{{\sf t}, k} }$.
Then, the unfolded sub-layer that mimics one MM iteration is constructed as
\begin{align} \label{q-learnable-update}
	\mbf{q}_k(\bsm{\Upsilon}_k) = \frac{1}{\eta_k}\left(\mbf{b}_k - \left(\mbf{N}_k -\eta_k\bsm{\Upsilon}_k\right)\underline{\mbf{v}}_k\right).
\end{align}
We assume the same number $I$ of MM sub-layers for different $k$.
At the $\ell$-th layer, let $\bsm{\Upsilon}_k^{\ell, i} \in \mathbb{C}^{D_k N_{{\sf t}, k} \times D_k N_{{\sf t}, k} }$ be the learnable parameters associated with the $i$-th MM sub-layer of $\mbf{v}_k$. 
The network architecture for $\{\mbf{V}_k\}_{k\in\mathcal{K}}$ update at the $\ell$-th layer is given as follows:
\begin{mdframed}[hidealllines=true, backgroundcolor=black!20]
	\vspace{-.6em}
	\begin{subequations}\label{Vv-MM-learnable-update}
		\begin{align} 
			&\textbf{for}~ k = 1, \dots, K \textbf{~do} \nonumber \\ 
			&~~~\textbf{for}~ i =1, \dots, I \textbf{~do} \nonumber \\
			&~~~~~~\mbf{q}_k^{\ell, i} \; = \frac{1}{\eta_k^\ell} \left(\mbf{b}_k^\ell - \left(\mbf{N}_k^\ell -\eta_k^\ell\bsm{\Upsilon}_k^{\ell, i}\right)\mbf{v}_k^{\ell, i-1}\right); \\
			&~~~~~~\mbf{v}_k^{\ell, i} \; = \mbf{q}_k^{\ell, i} \min \left\{ \frac{\sqrt{P_k}}{\big\| \mbf{q}_k^{\ell, i} \big\|_2} , 1 \right\}; \label{v-MM-update} \\
			&~~~\textbf{end} \nonumber \\
			&~~~\mbf{V}_k^\ell \, \, \, = \mathrm{vec}^{-1} \left( \mbf{D}_k^{-1}  \mbf{v}_k^{\ell, I} \right); \label{V-MM-update} \\
			&\textbf{end} \nonumber 
		\end{align}
	\end{subequations}
\end{mdframed}
We use 
$\bsm{\Upsilon}^\ell \triangleq \{\bsm{\Upsilon}^\ell_k\}_{k\in\mathcal{K}}$ to collect the learnable parameters, where $\bsm{\Upsilon}_k^{\ell} \triangleq \{\bsm{\Upsilon}_k^{\ell,i}\}_{i \in \mathcal{I}}$ with $\mathcal{I} \triangleq \{1, \dots, I\}$.
As illustrated in Fig. \ref{proposed-DU-BCA-MM}, the enhanced DU-BCA-MM precoder incorporates \eqref{U-learnable-update}, \eqref{W-learnable-update}, and \eqref{Vv-MM-learnable-update} to construct one layer of the precoding network.
Table \ref{parameters} summarizes the learnable parameters introduced in the vanilla DU-BCA precoder and the enhanced DU-BCA-MM precoder.

\begin{table*}[t]
	\caption{Summary of the Learnable Parameters in Vanilla DU-BCA Precoder and Enhanced DU-BCA-MM Precoder}
	\label{parameters}
	\centering
		\begin{tabular}{c|cc}
			\toprule
			Precoder architecture & Learnable parameters (at the $\ell$-th layer)  & Number of learnable parameters (per layer) \\
			\midrule
			Vanilla DU-BCA & $\bsm{\Theta}^\ell ,\bsm{\Phi}^\ell, \bsm{\Psi}^\ell, \bsm{\Omega}^\ell, \bsm{\lambda}^\ell$ & $2 N_{\sf r}^2 + D^2 + \sum_{k \in \mathcal{K}}D_k^2 N_{{\sf t}, k}^2 + K$ \\
			Enhanced DU-BCA-MM & $\bsm{\Theta}^\ell ,\bsm{\Phi}^\ell, \bsm{\Psi}^\ell, \bsm{\Upsilon}^\ell$ & $2 N_{\sf r}^2 + D^2 + \sum_{k \in \mathcal{K}}ID_k^2 N_{{\sf t}, k}^2 $  \\
			\bottomrule
		\end{tabular}
\end{table*}

\begin{table*}[t]
	\caption{Implementation Details of the Baselines}
	\label{baselines}
	\centering
	\begin{tabular}{cc|cccc}
		\toprule
		Figure & Scheme & Feature encoder & Precoder & Classifier & Training strategy$^*$ \\
		\midrule
		\multirow{2}{*}{Fig. \ref{Work3_Fig1b}} & Proposed framework & \multirow{2}{*}{VGG11} & \multirow{2}{*}{6-layer DU-BCA-MM} & \multirow{2}{*}{MAP} & E2E with FE\&P-PT \\ 
		& E2E learning w/o pretraining & & & & E2E w/o FE\&P-PT \\
		\midrule
		\multirow{5}{*}{Fig. \ref{Work3_Fig2}} & Vanilla DU-BCA & \multirow{5}{*}{VGG11} & DU-BCA & \multirow{5}{*}{MAP} &  E2E with FE\&P-PT \\
		& Enhanced DU-BCA-MM & & DU-BCA-MM & & E2E with FE\&P-PT \\
		& BCA algorithm & & BCA algorithm & & FE-PT only \\
		& BCA-MM algorithm & & BCA-MM algorithm & & FE-PT only \\
		& Black-box precoding network & & ResNet18 & & E2E with FE\&P-PT \\
		\midrule
		\multirow{2}{*}{Fig. \ref{Work3_Fig4}} & Proposed framework & \multirow{2}{*}{VGG11}  & 6-layer DU-BCA-MM & MAP & E2E with FE\&P-PT \\
		& ST-FE\&C with LMMSE TRx & & LMMSE algorithm & LMMSE detector with MLP & ST-FE\&C  \\
		\midrule
		\multirow{2}{*}{Fig. \ref{FigR1}} & Proposed framework & \multirow{2}{*}{ResNet18} & 6-layer DU-BCA-MM &  MAP & E2E with FE\&P-PT \\
		& E2E-FE\&C with LMMSE TRx & &  LMMSE algorithm &  LMMSE detector with MLP &  E2E-FE\&C \\
		\midrule
		Fig. \ref{Work3_Fig3}  & All schemes & VGG11 & 6-layer DU-BCA-MM & MAP &  E2E with FE\&P-PT \\
		\bottomrule
	\end{tabular}\\
	~\\
	\rightline{$^*$\textbf{E2E with FE\&P-PT:} E2E learning with feature encoding and precoding pretraining;
		\textbf{E2E w/o FE\&P-PT:} E2E learning without feature encoding and~}
	\leftline{precoding pretraining; \textbf{FE-PT only:} feature encoding pretraining only; \textbf{ST-FE\&C:} separately trained feature encoding and classification; \textbf{E2E-FE\&C:} E2E}
	\leftline{trained feature encoding and classification. }
\end{table*}

\section{Simulation Results} \label{Sec_simulation}
\subsection{Experimental Setup}
\subsubsection{Datasets}
We conduct the classification task on the CIFAR-10 \cite{krizhevsky2009learning} and ModelNet10 \cite{wu20153d} datasets.
The CIFAR-10 dataset consists of $10$ classes of color images.
The ModelNet10 dataset is a multi-view image dataset, which contains $10$ classes of computer-aided design (CAD) objects (e.g., sofa, bathtub, bed).
Each object in ModelNet10 is captured from twelve distinct views.

\subsubsection{System and Communication Settings}
We assume a single device, i.e., $K=1$, for the experiments on CIFAR-10.
For the experiments on ModelNet10, a total of $K=3$ devices is considered, with the input views selected among the twelve available views. 
Unless specified otherwise, we adopt the following default settings:
$D = 8$, $N_{{\sf t}} = 8$, and $N_{\sf r} = 8$ for the experiments on CIFAR-10;
$D_k = 4$, $N_{{\sf t}, k} = 4$, and $N_{\sf r} = 8$ for the experiments on ModelNet10.
We assume the same distance $d = 80$ m from each edge device to the server, and the path loss is set as $32.6 + 36.7 \lg d$ dB.
The channel between each edge device and the server is independent and modeled by Rician fading as
\begin{align}
	\mbf{H}_k = \sqrt{\frac{\kappa}{\kappa + 1}} \mbf{H}_k^{\mathrm{LoS}} + \sqrt{\frac{1}{\kappa + 1}} \mbf{H}_k^{\mathrm{NLoS}},
\end{align}
where $\mbf{H}_k^{\mathrm{LoS}}$ is the line-of-sight (LoS) component, $\mbf{H}_k^{\mathrm{NLoS}} \sim \mathcal{CN}(\mbf{0}, \mbf{I})$ is the non-LoS (NLoS) component, and the Rician factor is set to $\kappa = 1$.
The power budget at each device is set equally as $P_k = P_0$.
Unless specified otherwise, we fix the noise variance $\sigma^2 = - 80$ dBm.

\subsubsection{Neural Network Architecture and Learning Configurations}
For the experiments on CIFAR-10, we adopt the ResNet18 \cite{he2016deep} as the backbone of the feature encoder.
For the experiments on ModelNet10, each device employs a VGG11 \cite{simonyan2014very} as the feature encoder backbone.
Two fully connected layers are appended to the output of ResNet18/VGG11 for feature dimension reduction. 
We treat the first and second halves of the output layer as the real and imaginary parts of the feature vector, respectively.
For both the vanilla DU-BCA precoder and the enhanced DU-BCA-MM precoder, we assume $L=6$ unfolded layers unless specified otherwise.
Each layer of the enhanced DU-BCA-MM precoder adopts $I=2$ MM sub-layers for each $\mbf{v}_k$.
Both precoders are required to deal with complex numbers, which are not supported by current deep learning platforms.
To resolve this issue, we use PyTorch to manually implement the required complex operations by their corresponding real-valued representations \cite{hu2021joint}.
For example, we implement the complex matrix multiplication by separately computing its real and imaginary parts as
\begin{align}
	\mathrm{Re} \{\mbf{A}\mbf{B}\} &= \mathrm{Re} \{\mbf{A}\} \mathrm{Re} \{\mbf{B}\} - \mathrm{Im} \{\mbf{A}\} \mathrm{Im} \{\mbf{B}\}, \\
	\mathrm{Im} \{\mbf{A}\mbf{B}\} &= \mathrm{Re} \{\mbf{A}\} \mathrm{Im} \{\mbf{B}\} + \mathrm{Im} \{\mbf{A}\} \mathrm{Re} \{\mbf{B}\}.
\end{align}
We adopt the Adam optimizer in all simulations.
The learning rates in Algorithms \ref{alg1}, \ref{alg2}, and \ref{alg3} are set to $1 \times 10^{-4}$, $1 \times 10^{-1}$, and $1\times 10^{-4}$, respectively.
The corresponding batch sizes are chosen as $B_1 = 1000$, $B_2 = 200$, and $B_3 = 200$.
Moreover, $\epsilon^2$ is set to $0.5$ for feature encoding and $1 \times 10^{-6}$ for MIMO precoding.
All experiments are conducted on a NVIDIA RTX A6000 GPU and Intel Xeon w9-3475X CPU @ 2.20 GHz. 

\subsection{Performance Comparisons}

\begin{figure}
	[t]
	\centering
	\includegraphics[width=.9\columnwidth]{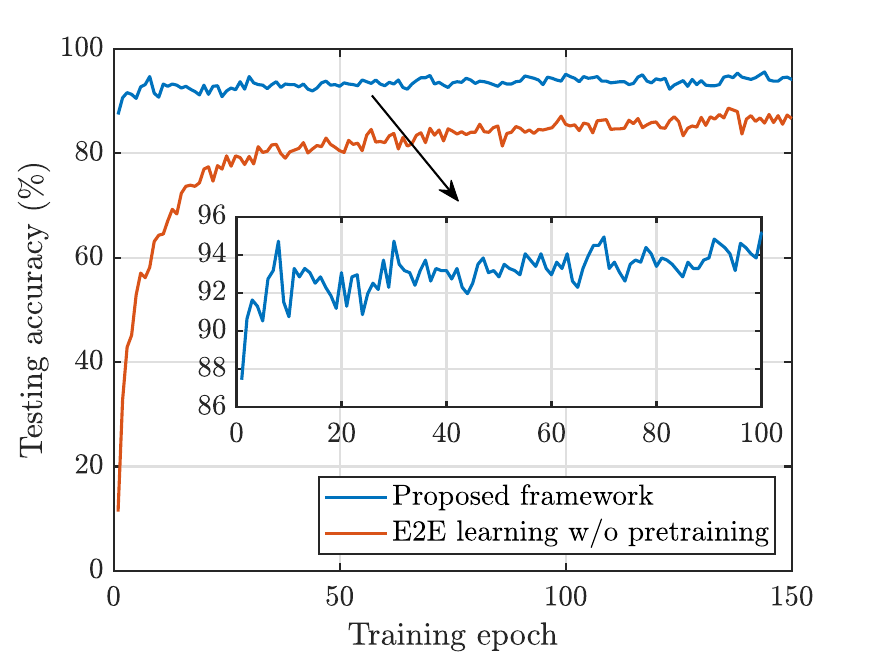}
	\caption{Testing accuracy at each training epoch in the E2E learning phase, where the power budget $P_0 = 15$ dBm and the number of transmit time slot $O=1$.
		The experiments are carried out on ModelNet10.
}
	\label{Work3_Fig1b}
\end{figure}

\begin{table}[t]
	\caption{Training Time of the Proposed Method During Decoupled Pretraining and E2E Fine-Tuning on ModelNet10}
	\label{training_consumption}
	\centering
		\begin{tabular}{c|cccc}
			\toprule
			& $D_k$ & $N_{\mathsf{t}, k}$  & $N_{\sf r}$&  \makecell{Training time \\ per epoch (s)} \\
			\midrule
			\multirow{4}{*}{\makecell{Feature encoder\\ (VGG11) \\ pretraining}} & $4$ & - &  -  & $14.17$ \\
			& $8$ & - & -   & $14.33$ \\
			& $12$ & - & -  & $14.53$ \\
			& $16$ & - & -  & $14.68$ \\
			\midrule
			\multirow{7}{*}{\makecell{Precoder\\ (DU-BCA-MM) \\ pretraining}} & $4$ & $4$ & $8$ &  $49.31$ \\
			& $4$ & $8$ & $8$ &  $49.17$ \\
			& $4$ & $12$ & $8$ &  $49.91$ \\
			& $4$ & $16$ & $8$ &  $51.86$ \\
			& $4$ & $4$ & $12$ &  $49.92$ \\
			& $4$ & $4$ & $16$ &  $52.70$ \\
			& $4$ & $4$ & $20$ &  $54.63$ \\
			\midrule 
			\multirow{7}{*}{\makecell{E2E fine-tuning}} & $4$ & $4$ & $8$  & $301.07$ \\
			& $4$ & $8$ & $8$ &  $303.24$ \\
			& $4$ & $12$ & $8$ &  $305.17$ \\
			& $4$ & $16$ & $8$ &  $310.13$ \\
			& $4$ & $4$ & $12$ &  $303.89$ \\
			& $4$ & $4$ & $16$ &  $308.48$ \\
			& $4$ & $4$ & $20$ & $317.62$ \\
			\bottomrule
		\end{tabular}
	\end{table}

\subsubsection{Indispensable Role of Decoupled Pretraining}
We validate the effectiveness of the proposed framework by comparing it with the E2E learning baseline without decoupled pretraining.
This baseline randomly initializes the network parameters and directly trains the feature encoder and precoder in an E2E manner.
It adopts the same MAP classifier and E2E cross-entropy loss \eqref{ce_empirical} as in the proposed method.\footnote{The implementation details of the baselines presented in each simulation figure are summarized in Table \ref{baselines}.}
Fig. \ref{Work3_Fig1b} illustrates the testing accuracy of the two schemes plotted against the E2E training epoch on ModelNet10.
It is seen that without proper initialization, the E2E learning experiences a slow convergence speed.
In contrast, our proposed framework, which leverages the network parameters learned in the decoupled pretraining phase for initialization, benefits from a ``warm start''.
We further harness an accuracy gain of around $6\%$ (from $88\%$ to $94\%$) within $10$ E2E training epochs.
In short, our proposed framework significantly improves the training efficiency and exhibits a better classification accuracy.
The reasons for the noticeable performance gap even after convergence are analyzed as follows.
The cross-entropy loss used in E2E learning is highly non-convex with respect to the NN parameters due to the underlying parameterization strategy.
As a result,  E2E training is prone to getting stuck in local optima if not accompanied by proper initialization or other effective strategies. 
On the other hand, by penalizing the features with a GM distribution, the decoupled pretraining provides useful model information on the learned features, which significantly reduces the training difficulty.
Moreover, it is theoretically established in \cite{global_geometric2024icml} that the MCR$^2$ objective, derived from the GM feature assumption, has a benign global optimization landscape.
Such a favorable landscape justifies why MCR$^2$ can be optimized well using simple learning algorithms such as gradient-based methods.

Table \ref{training_consumption} provides the training consumption during decoupled pretraining and E2E fine-tuning for various values of $D_k$, $N_{{\sf t}, k}$, and $N_{\sf r}$.
Other parameters are configured identically to those in Fig. \ref{Work3_Fig1b}.
Its shown that the pretraining requires notably less training time per epoch compared to the E2E fine-tuning, showcasing the feasibility and efficiency of the proposed framework.
In practice, we pretrain the feature encoder and the precoder for $400$ and $100$ epochs, respectively.
Then, the network undergoes E2E fine-tuning for another $10$ to $20$ epochs.
Based on the above, it can be easily verified that our proposed framework saves the training overhead by orders of magnitude.

\begin{figure}
	[t]
	\centering
	\subfigure[$\Delta \widetilde{\mathcal{R}}(R; Y|S)$ vs. number of layers/iterations] 
	{\includegraphics[width=.9\columnwidth]{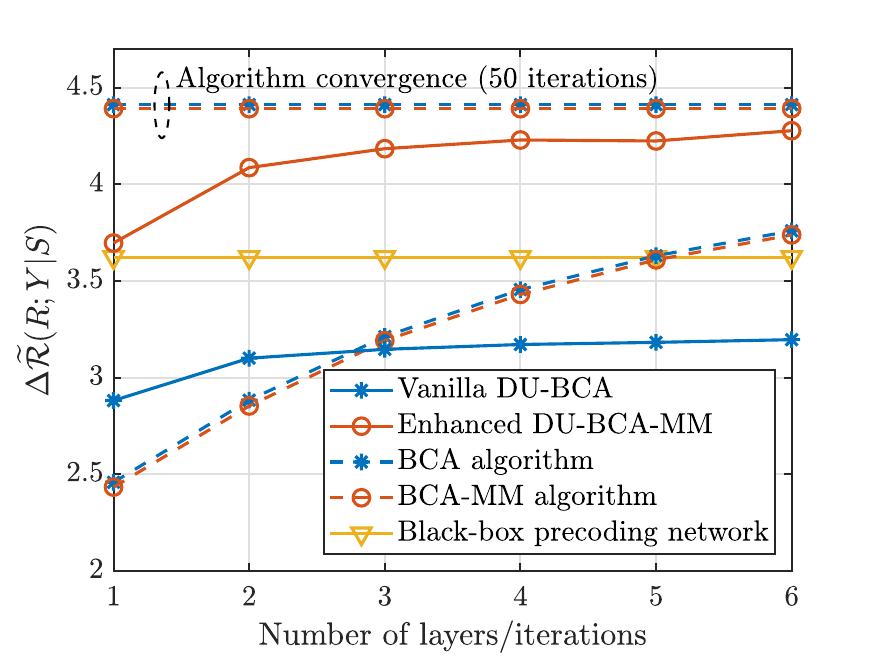}\label{Work3_Fig2a}}
	\subfigure[E2E Inference accuracy vs. number of layers/iterations]
	{\includegraphics[width=.9\columnwidth]{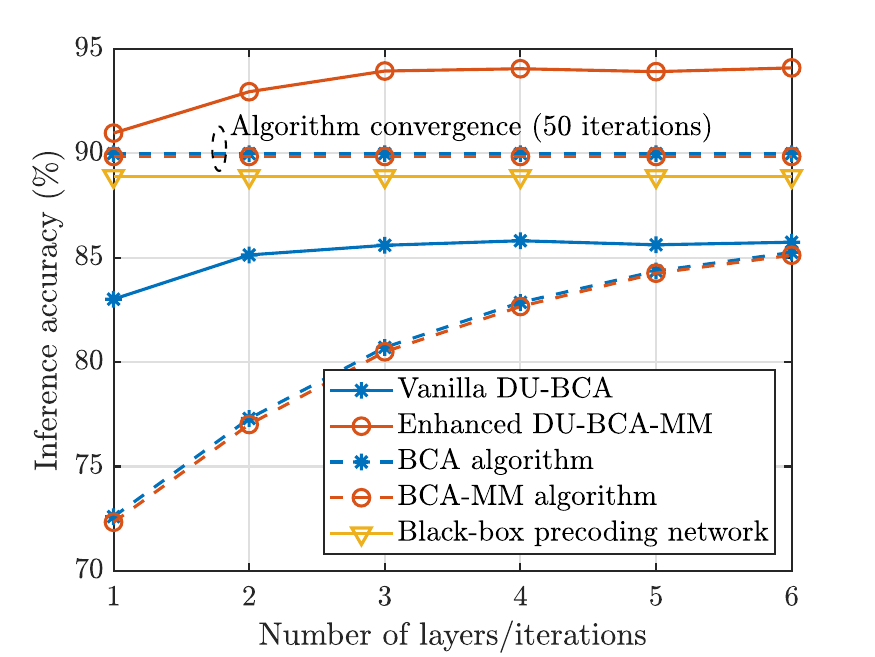}\label{Work3_Fig2b}}
	\caption{Performance comparisons of different precoding schemes on ModelNet10 for a varying number of unfolded layers/algorithm iterations, where $P_0 = 15$ dBm and $O=1$.
		We run the BCA and BCA-MM algorithms for $50$ iterations to obtained the converged performance.}
	\label{Work3_Fig2}
\end{figure}

\subsubsection{Performance and Complexity Comparisons of Different Precoders}
We consider both the black-box network and optimization-based implementations of MCR$^2$ precoding for comparison.
For the black-box precoding network, we utilize the ResNet18 \cite{he2016deep} architecture with essential modifications made to accommodate the precoding problem. 
The ResNet18 takes the real and imaginary parts of the channel matrix $\mbf{H}$ as input.
The output of the network is $\{\mbf{V}_k\}_{k \in \mathcal{K}}$, followed by the projection operator \eqref{projection} to ensure the satisfaction of the power constraints.
For optimization-based precoding, this category of baselines leverages the BCA \cite{cai2023multi} or the BCA-MM algorithms to solve the MCR$^2$ precoding problem instead an NN, as described in Sections \ref{Sec-BCA-Algorithm} and \ref{Sec-BCA-MM}, respectively.
Focusing on the precoding pretraining, Fig. \ref{Work3_Fig2a} compares the achieved MCR$^2$ values of the deep unfolding networks with their optimization algorithm counterparts on ModelNet10.
The result of the black-box precoding network is also provided.
It is shown that for the same number of unfolded layers/algorithm iterations, the enhanced DU-BCA-MM precoder significantly outperforms its base BCA-MM algorithm.
The MCR$^2$ value improvement is around $52\%$ and $15\%$ for $1$ and $6$ layer(s)/iteration(s), respectively.
We speculate that the enhanced DU-BCA-MM precoder learns a faster ascending trajectory compared to the BCA-MM algorithm operating in a block coordinate ascent manner.
Notably, the enhanced DU-BCA-MM precoder with only $6$ unfolded layers approaches the converged performance of the BCA and BCA-MM algorithms.
The enhanced DU-BCA-MM precoder also outperforms the black-box NN owing to its tailored architecture for the MCR$^2$ precoding problem. 
The vanilla DU-BCA precoder, as expected, does not perform well especially when scaling up to a slightly large number of layers.

Then, we employ the precoding networks pretrained in Fig. \ref{Work3_Fig2a} for E2E learning.
The inference accuracy results are reported in Fig. \ref{Work3_Fig2b}.
The baselines employing the BCA/BCA-MM algorithms are not fine-tuned in an E2E manner because no learnable parameters are introduced for precoding.
Remarkably, even though the enhanced DU-BCA-MM precoder in Fig. \ref{Work3_Fig2a} does not achieve a larger MCR$^2$ value compared to the converged performance of the BCA/BCA-MM algorithm,
adopting the pretrained DU-BCA-MM precoder for E2E learning can finally outperform all these benchmarks.
This highlights the necessity of E2E learning to better align the objectives of feature encoding and precoding.
In addition, it is seen that approximately $3$ unfolded layers are adequate for the enhanced DU-BCA-MM precoder.
Further increasing the number of layers does not yield considerable improvements in classification accuracy.

\begin{table}[t]
	\caption{Execution Latency (s) in CPU Time of Different Precoding Schemes on ModelNet10}
	\label{deep_unfolding_execution_time}
	\centering
	\begin{tabular}{cccc}
		\toprule
		DU-BCA & BCA Alg. & DU-BCA-MM & BCA-MM Alg. \\
		\midrule
		$0.008$ & $0.046$ & $0.011$ & $0.053$  \\
		\bottomrule
	\end{tabular}
\end{table}

\begin{figure}
	[t]
	\centering
	\includegraphics[width=.9\columnwidth]{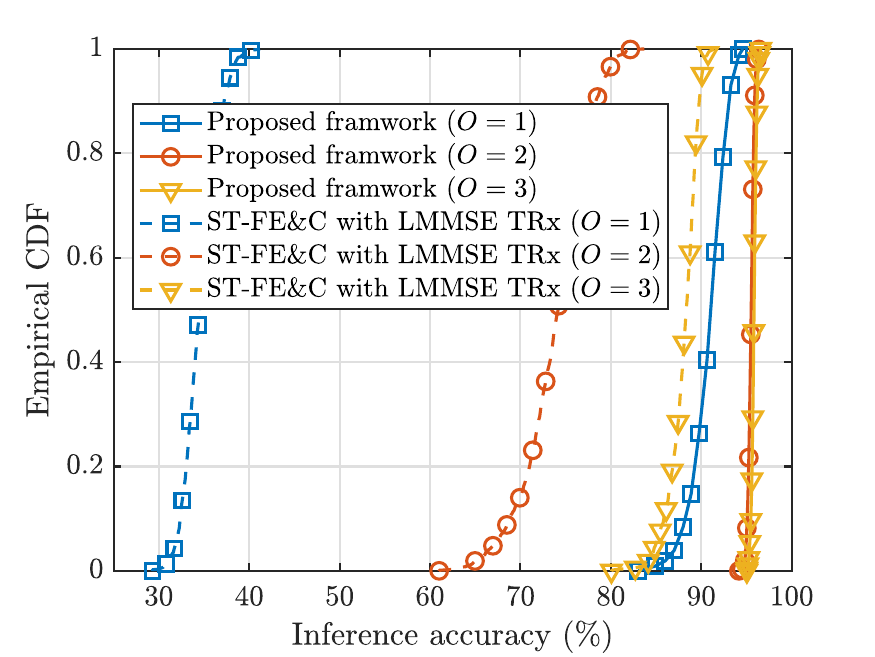}
	\caption{Empirical CDF of inference accuracy for random channel realizations, where $P_0 = 10$ dBm.
	The experiments are carried out on ModelNet10.}
	\label{Work3_Fig4}
\end{figure}

Table \ref{deep_unfolding_execution_time} presents the execution times of different precoding schemes on ModelNet10, all using the same number of $6$ unfolded layers or algorithm iterations.
All timings are performed on the CPU for consistency.
We see that the two deep unfolding networks require less CPU running time than their respective base algorithms, showcasing remarkably low complexity.
This indicates the possibility of incorporating these proposed networks into practical engineering applications.

\subsubsection{Comparisons with LMMSE-Based Feature Transmission Design}
We next investigate the impact of varying the number of transmit time slots on the inference accuracy.
In addition to our proposed framework, we consider an LMMSE benchmark that adopts the MCR$^2$ feature encoder pretrained in Algorithm \ref{alg1}, but applies the LMMSE precoder \cite[Appendix A]{cai2023multi} and the LMMSE detector \cite{zhijin2021multiuser} in the feature transmission pipeline.
The recovered features undergo classification via a multilayer perceptron (MLP) trained on noiseless features.
We refer to this baseline as ``separately trained feature encoding and classification (ST-FE\&C) with LMMSE transceiver (TRx)''.
In Fig. \ref{Work3_Fig4}, we generate $1000$ independent channel realizations for each testing sample in the ModelNet10 dataset, and plot the cumulative distribution function (CDF) of the inference accuracy. 
We assume that the channel remains unchanged if multiple transmit time slots are considered.
One can see that increasing the number of transmit time slots improves the accuracy.
For our proposed framework, a larger number of transmit time slots results in more stable performance, indicated by a smaller variance of the inference accuracy. 
The performance gap between the proposed framework and the LMMSE benchmark is extremely large when $O=1$.
In this case, the LMMSE detector fails since $\mbf{H} \mbf{V}$ is a fat matrix ($ON_{\sf r} < D$).
When two or three time slots are used for transmission, the performance gap is still non-negligible even with the resulting tall matrix $\mbf{H}\mbf{V}$ ($ON_{\sf r} > D$).

\begin{figure}[t]
	\centering
	\subfigure[Acc. vs. $N_{\sf r}$ ($O=1$)]
	{\includegraphics[width=.492\columnwidth]{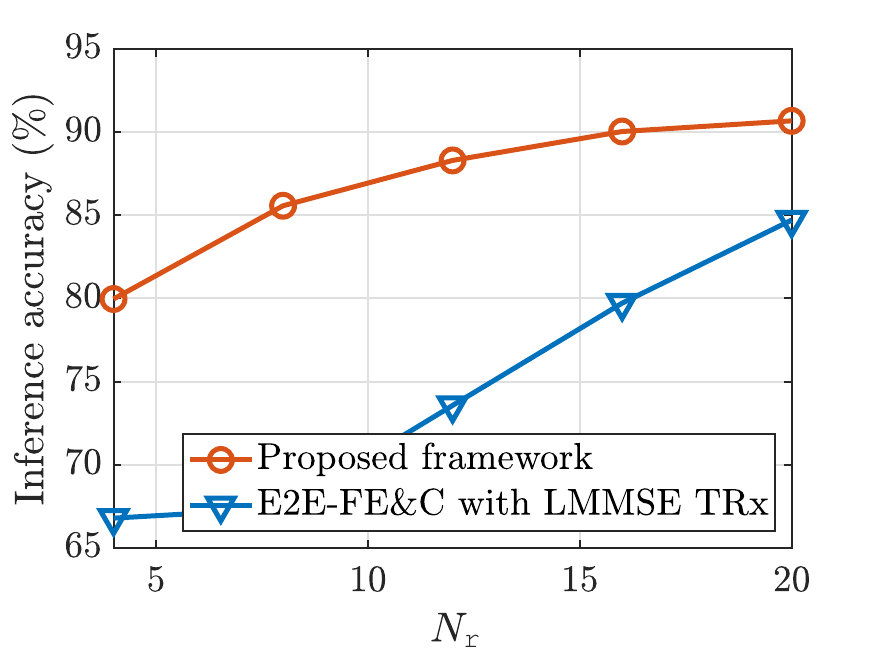}\label{Fig1a_ModelNet}}
	\subfigure[Acc. vs. $N_{\sf r}$ ($O=2$)]
	{\includegraphics[width=.492\columnwidth]{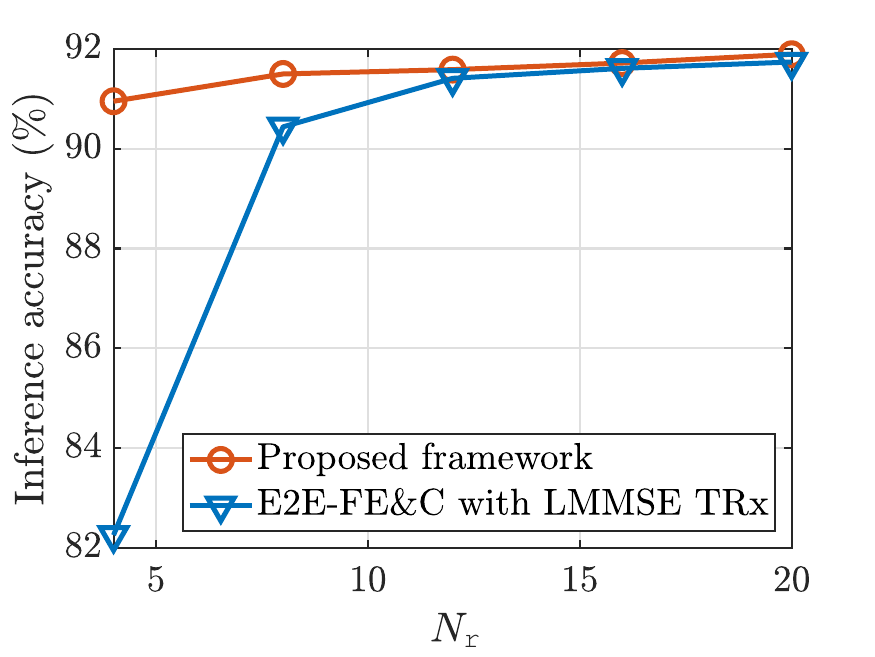}\label{Fig1b_ModelNet}}
	\subfigure[Acc. vs. $N_{\sf t}$ ($O=1$)]
	{\includegraphics[width=.492\columnwidth]{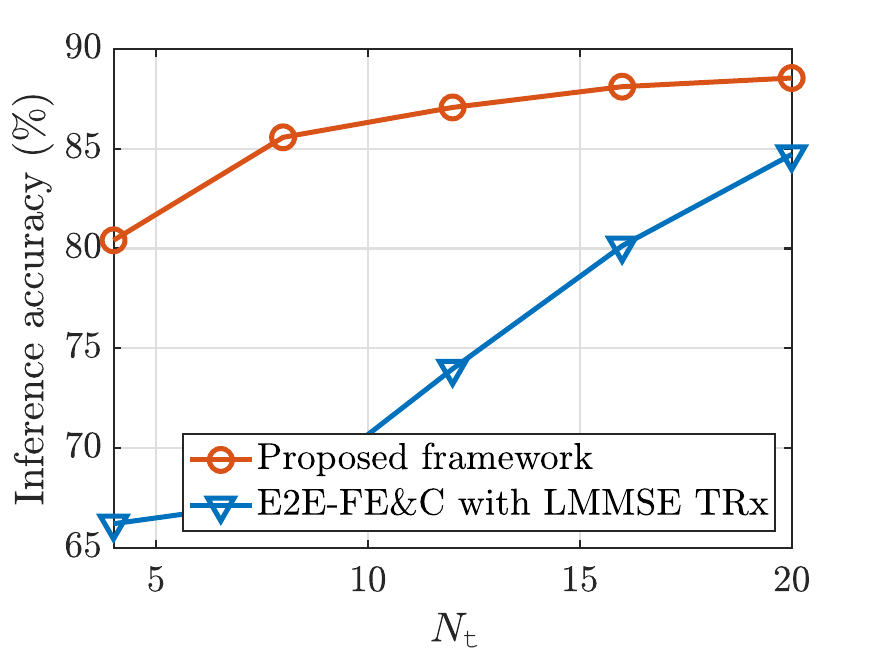}\label{Fig1a_ModelNet}}
	\subfigure[Acc. vs. $N_{\sf t}$ ($O=2$)]
	{\includegraphics[width=.492\columnwidth]{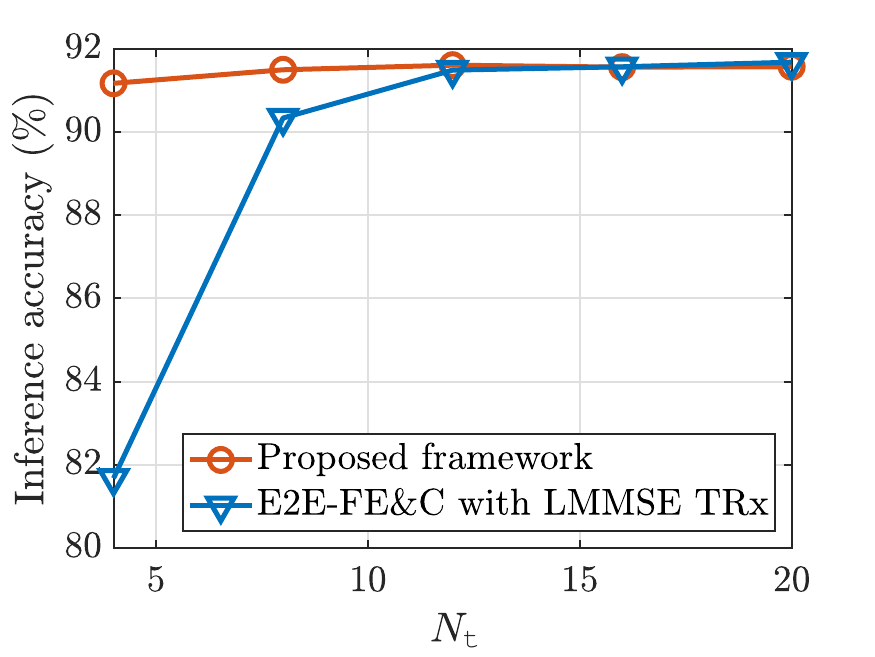}\label{Fig1b_ModelNet}}
	\caption{Inference accuracy vs. $N_{\sf r}$ and $N_{\sf t}$ for $O\in \left\{1, 2\right\}$, where $P_0=10$ dBm.
		The experiments are carried out on CIFAR-10.}
	\label{FigR1}
\end{figure}

We further explore the performance limit of LMMSE-based feature transmission schemes by E2E learning.
The refined LMMSE benchmark with E2E learning trains the feature encoder and the MLP classifier together using the cross-entropy loss, incorporating the LMMSE precoder \cite[Appendix A]{cai2023multi} and the LMMSE detector \cite{zhijin2021multiuser} for feature transmission and recovery during training.
We refer to this benchmark as ``E2E trained feature encoding and classification (E2E-FE\&C) with LMMSE TRx''.
Fig. \ref{FigR1} compares the inference accuracy of the proposed framework with the aforementioned benchmark on CIFAR-10 across different values of the transmit and receive antennas. 
By comparing Fig. \ref{Work3_Fig4} and Fig. \ref{FigR1}, it is evident that E2E learning improves the performance of the LMMSE-based feature transmission scheme.
This is because E2E learning allows the network, particularly the classifier, to better mitigate the impact of detection errors.
Nevertheless, the proposed framework still offers a substantial accuracy gain in most cases.
The performance gain primarily stems from the aligned objectives of learning and communication.
 
\begin{figure}
	[t]
	\centering
	\includegraphics[width=.9\columnwidth]{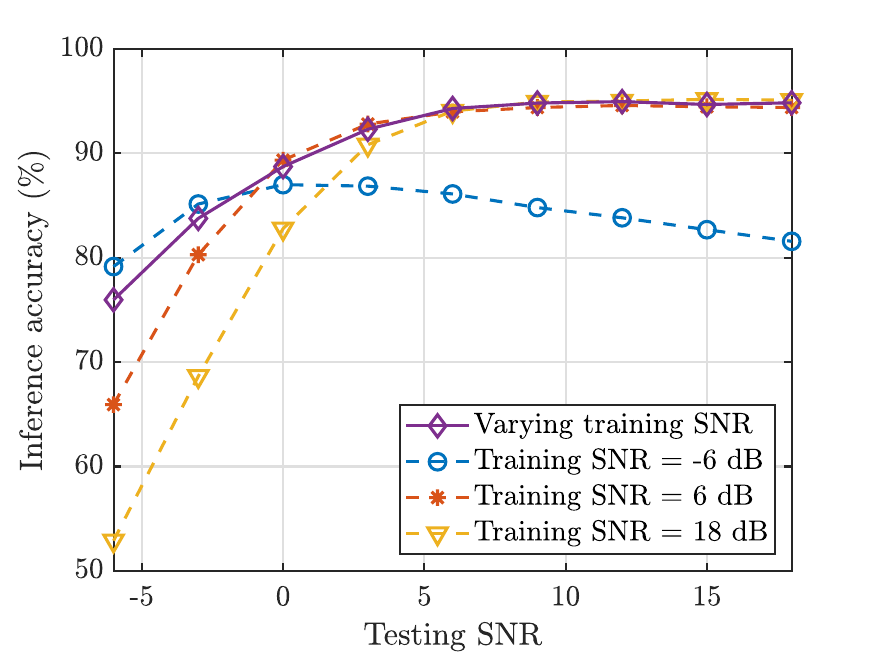}
	\caption{Inference accuracy over different testing SNRs using varying or fixed training SNRs, where $P_0 = 15$ dBm, $O=1$, and the noise variance $\sigma^2$ is chosen according to the SNR.
		The experiments are carried out on ModelNet10.}
	\label{Work3_Fig3}
\end{figure}

\subsubsection{Robustness and Generalization Capabilities}
We verify the robustness of the proposed method against SNR variations on ModelNet10.
We train the network using different SNRs in the set $\left\{-6 \text{~dB}, 0 \text{~dB}, 6 \text{~dB}, 12 \text{~dB}, 18 \text{~dB}\right\}$.
Following \cite{cg2020tsp}, we first train the network at high SNR in order to learn the intrinsic structure of the classification problem.
We gradually decrease the SNR in the subsequent training process to guarantee robustness.
Then, we use the SNR uniformly sampled from the set for training to further improve the performance.
It is shown in Fig. \ref{Work3_Fig3} that compared to the network trained at a specific SNR, the network trained with different SNRs adapts better to SNR variations.
Nevertheless, there is still a performance gap in the low SNR regime compared to the ideal case of employing a network specifically trained for low SNR ($-6$ dB). 
Some recent works attempt to fill this gap by introducing additional architectures such as attention modules \cite{wu2023deep} or Hypernetworks \cite{xie2024deep}, which is worth further investigation.

\begin{table}[t]
	\caption{Out-of-Distribution Performance of the Proposed Method on CIFAR-10 over Different Training and Testing Rician Factors}
	\label{OoD_Rician_factor}
	\centering
	\begin{tabular}{c|ccccc}
		\toprule
		& $\!\!\kappa_{\text{test}} \!=\! 0.1$ & $\!\!\kappa_{\text{test}} \!= \!0.5$ & $\!\kappa_{\text{test}} \!=\!1$ & $\!\kappa_{\text{test}} \!=\!5$ & $\kappa_{\text{test}} \!=\!10$  \\
		\midrule 
		$\!\kappa_{\text{train}} \!=\! 0.1$ & $0.9016$ & $0.8969$ & $0.8895$ & $0.8520$ & $0.8343$ \\
		$\!\kappa_{\text{train}} \!=\! 0.5$ & $0.8961$ & $0.9134$ & $0.9168$ & $0.9157$ & $0.9142$  \\
		$\!\kappa_{\text{train}} \!=\! 1$ & $0.8864$ & $0.9104$ & $0.9174$ & $0.9186$ & $0.9180$  \\
		$\!\kappa_{\text{train}} \!=\! 5$ & $0.8368$ & $0.8739$ & $0.9002$ & $0.9186$ & $0.9196$  \\
		$\!\kappa_{\text{train}} \!=\! 10$ & $0.8222$ & $0.8628$ & $0.8915$ & $0.9172$ & $0.9188$  \\
		\bottomrule
	\end{tabular}
\end{table}

Table \ref{OoD_Rician_factor} presents the out-of-distribution generalization performance of the proposed framework under channel distribution shifts.
In each row of Table \ref{OoD_Rician_factor}, the network is trained on a specific Rician factor and tested across varying Rician factors.
The results show that the performance drop is negligible with a moderate change in the channel distribution.
The most extreme case occurs when training at $\kappa_{\text{train}} = 10$ but testing at $\kappa_{\text{test}} = 0.1$, resulting in an $11.75\%$ performance drop.
Overall, Table \ref{OoD_Rician_factor} showcases the strong generalization capabilities of the proposed framework across varying Rician factors.

\section{Conclusion and Discussion}\label{Sec_Conclusions}
In this paper, we studied the E2E learning design of multi-device cooperative edge inference over a wireless MIMO multiple access channel.
We formulated the joint design of feature encoding, precoding, and classification as an E2E conditional mutual information maximization problem.
To reduce the training cost, we established a decoupled pretraining framework that exploits the close connection between mutual information and coding rate reduction.
Owing to the aligned objectives of each individual component, the decoupled pretraining substantially reduces the E2E learning overhead.
Simulation results validated the superior classification accuracy of our approach compared to various baselines. 

The proposed framework offers promising opportunities for generalization to a variety of task-oriented applications. 
The conditional mutual information maximization considered in this paper stands for a universal objective for different tasks.
However, although the objective can be approximated in closed-form for the classification task, it is in general intractable for other machine learning tasks due to the high dimensional integrals with even unknown distributions. 
To tackle this difficulty, one may resort to the variational approximations of mutual information \cite{poole2019variational} in order to obtain a tractable form amenable for learning and optimization.
On the other hand, the proposed framework may be applicable to other tasks through a direct generalization and improvement of the MCR$^2$ objective.
For example, \cite{yu2023white} and \cite{pai2023masked} add a sparsity-inducing term on the rate reduction, resulting in the sparse rate reduction objective.
This objective serves as the criterion for network construction, which performs well on a variety of tasks such as autoencoding, image completion, language understanding, and text generation.

\bibliographystyle{IEEEtran}
\bibliography{IEEEabrv,mybib}

\end{document}